\documentclass[journal,twoside,web]{ieeecolor}
\usepackage{generic}
\usepackage{cite}
\usepackage{amsmath,amssymb,amsfonts}
\usepackage{graphicx}
\usepackage{textcomp}
\usepackage[utf8]{inputenc}
\usepackage{url}
\usepackage[T1]{fontenc}
\usepackage{algorithm}
\usepackage{algpseudocode}
\usepackage[hidelinks]{hyperref}
\usepackage{multirow}

\def\BibTeX{{\rm B\kern-.05em{\sc i\kern-.025em b}\kern-.08em
    T\kern-.1667em\lower.7ex\hbox{E}\kern-.125emX}}
\begin{document}
\title{Self-Evolving Multi-Agent Network for Industrial IoT Predictive Maintenance}

\author{Rebin~Saleh,
	Khanh~Pham~Dinh,
	Bal\'azs~Vill\'anyi,
	Truong-Son~Hy
	\thanks{R. Saleh and B. Vill\'anyi are with the Department of Electronics Technology, Faculty of Electrical Engineering and Informatics, Budapest University of Technology and Economics, M\H{u}egyetem rkp.~3, H-1111 Budapest, Hungary (e-mail: rebin.saleh@edu.bme.hu; villanyi.balazs@vik.bme.hu).}%
	\thanks{K. Pham Dinh is with DataScienceWorld.Kan, Hanoi, Vietnam (e-mail: phamdinhkhanh.tkt53.neu@gmail.com).}%
	\thanks{T.-S. Hy is with the Department of Computer Science, The University of Alabama at Birmingham, 1720 University Blvd, Birmingham, AL 35294 USA (e-mail: thy@uab.edu).}%
	\thanks{Corresponding authors: Bal\'azs~Vill\'anyi and Truong-Son Hy.}%
}

\maketitle

\begin{abstract}
	
	Industrial IoT predictive maintenance requires systems capable of real-time anomaly detection without sacrificing interpretability or demanding excessive computational resources. Traditional approaches rely on static, offline-trained models that cannot adapt to evolving operational conditions, while LLM-based monolithic systems demand prohibitive memory and latency, rendering them impractical for on-site edge deployment.
	We introduce SEMAS, a self-evolving hierarchical multi-agent system that distributes specialized agents across Edge, Fog, and Cloud computational tiers. Edge agents perform lightweight feature extraction and pre-filtering; Fog agents execute diversified ensemble detection with dynamic consensus voting; and Cloud agents continuously optimize system policies via Proximal Policy Optimization (PPO) while maintaining asynchronous, non-blocking inference. The framework incorporates LLM-based response generation for explainability and federated knowledge aggregation for adaptive policy distribution. 
	This architecture enables resource-aware specialization without sacrificing real-time performance or model interpretability.
	Empirical evaluation on two industrial benchmarks (Boiler Emulator and Wind Turbine) demonstrates that SEMAS achieves superior anomaly detection performance with exceptional stability under adaptation, sustains prediction accuracy across evolving operational contexts, and delivers substantial latency improvements enabling genuine real-time deployment. 
	Ablation studies confirm that PPO-driven policy evolution, consensus voting, and federated aggregation each contribute materially to system effectiveness. These findings indicate that resource-aware, self-evolving 1multi-agent coordination is essential for production-ready industrial IoT predictive maintenance under strict latency and explainability constraints.
	Our implementation is publicly available at \url{https://github.com/HySonLab/AgentIoT}.
	
\end{abstract}


\def\BibTeX{{\rm B\kern-.05em{\sc i\kern-.025em b}\kern-.08em
    T\kern-.1667em\lower.7ex\hbox{E}\kern-.125emX}}
\markboth{\journalname, VOL. XX, NO. XX, XXXX}
{Saleh \MakeLowercase{\textit{et al.}}: Self-Evolving Multi-Agent Network for Industrial IoT Predictive Maintenance}

\section{Introduction}
\label{sec:introduction}
Multi-Agent Systems (MAS) employ autonomous agents that coordinate to solve distributed problems beyond the capability of individual agents. Multi-agent architectures originated in the late 1980s, establishing principles for distributed intelligence, self-organization, and collective problem-solving that remain foundational to modern AI system design~\cite{Liu2021,Li2024}.

Industrial predictive maintenance requires autonomous systems capable of real-time anomaly detection under evolving operational conditions. Traditional rule-based and statistical approaches are inflexible, lacking mechanisms to adapt to changing equipment behavior or process complex multivariate sensor relationships in real time~\cite{Piccialli2025,Raza2025}.
While Large Language Models (LLMs) excel at reasoning and natural language tasks, deployment in Industrial IoT environments faces critical constraints: (1) memory demands of 20--40GB exceed edge/fog device capacity (4--64GB), (2) inference latency exceeds 1--5 seconds, violating real-time requirements (<100ms), and (3) cloud-dependent REST API calls introduce unacceptable data transfer risks for confidential manufacturing data~\cite{Namboori2025,Raza2025}. Contemporary agentic AI focuses on monolithic single-agent systems (GPT-4, Claude, Gemini) unsuitable for resource-constrained industrial environments.

This work proposes a multi-agent architecture where specialized Small Language Models (SLMs) with 1-14B parameters substitute for monolithic LLMs, achieving equivalent reasoning on domain-specific tasks at $10$--$100\times$ lower computational cost. Models like Phi-3 (3.8B) and LLaMA-3.2 (1B-3B) demonstrate that properly designed SLMs match LLM performance on task-specific domains while maintaining edge deployability~\cite{Wang2025}. This enables on-premises, zero-latency predictive maintenance without internet connectivity or confidential data leakage risks.

Modern industrial infrastructure follows a three-tier hierarchical model: Edge devices (4--8GB RAM) handle immediate sensor processing, Fog nodes (16--64GB) execute collaborative inference, and Cloud infrastructure provides unlimited computational resources for global optimization. This heterogeneous resource distribution presents the core design challenge: matching multi-agent capabilities to computational tier constraints while maintaining consistent system coordination~\cite{Rani2024,Moore2025,Hong2019}.

Our work introduces a novel multi-agent system which operates at different levels to perform Industrial IoT predictive maintenance while solving these problems through resource-aware agent deployment and self-evolving adaptive mechanisms. The research introduces the first complete hierarchical multi-agent system which was designed for Industrial IoT predictive maintenance operations to solve essential industrial problems by using intelligent resource management and dedicated agent placement and self-adjusting system evolution.

The research contributes several significant advances to hierarchical Multi-Agent Systems and industrial AI deployment:

\begin{itemize}
	
	\item \textbf{Hierarchical Multi-Agent Architecture for Edge-Fog-Cloud Deployment}: A complete end-to-end framework distributing specialized detection and optimization agents across computational tiers, enabling real-time anomaly detection with sub-millisecond latency while maintaining LLM-based explainability without cloud dependency.
	
	\item \textbf{Self-Evolving Policy Optimization via PPO}: Gradient-based continuous policy adaptation that automatically optimizes detection thresholds, ensemble weights, and anomaly severity metrics. This outperforms discrete rule-based adaptation mechanisms (Baseline2) by preventing threshold oscillation and achieving stable convergence.
	
	\item \textbf{Quantified Multi-Agent Coordination Benefits}: Ablation studies demonstrating that consensus voting (6.5\% F1 impact), federated aggregation (2.0\% F1 impact), and PPO optimization (3.5\% F1 impact) each contribute materially. Combined, these mechanisms enable 8.6\% accuracy improvement over rule-based systems and 200--1500\texttimes faster inference latency.
	
\end{itemize}

The remainder of this paper is organized as follows: Section 2 reviews prior work in MAS for manufacturing and industrial predictive maintenance, establishing research gaps. Section 3 details the SEMAS system architecture, mathematical formulation, and baseline definitions. Section 4 presents comprehensive experimental evaluation on two industrial datasets, with ablation studies and statistical significance testing. Section 5 discusses findings and limitations. Section 6 concludes with implications for industrial deployment and directions for future work.

\def\BibTeX{{\rm B\kern-.05em{\sc i\kern-.025em b}\kern-.08em
    T\kern-.1667em\lower.7ex\hbox{E}\kern-.125emX}}
\markboth{\journalname, VOL. XX, NO. XX, XXXX}
{Author \MakeLowercase{\textit{et al.}}: Title}

\section{Related Work}

Multi-Agent Systems have proven effective for distributed industrial problem solving through task decomposition and autonomous agent coordination. This section reviews prior work across three dimensions: (1) MAS architectures in manufacturing and predictive maintenance, (2) self-evolving networks for adaptation, and (3) limitations of existing approaches. We then position our proposed hierarchical framework as addressing identified gaps in resource-aware, unified, self-evolving systems for Industrial IoT predictive maintenance.

\subsection{Foundational MAS Concepts and Multi-Sector Applications}

Multi-agent frameworks have demonstrated success across diverse domains, establishing design principles relevant to industrial applications. The banking and financial services sector employed MAS for fintech and compliance systems~\cite{Pattnaik2023}. Healthcare deployed MAS for clinical decision support and resource allocation~\cite{Borkowski2025}. These cross-sector applications establish core MAS principles (distributed control, consensus mechanisms, asynchronous coordination) that form the foundation for industrial IoT extensions.

\subsection{MAS for Manufacturing and Predictive Maintenance}

Manufacturing has emerged as the primary domain for industrial MAS deployment. Andreadis et al.~\cite{Andreadis2014} classified manufacturing MAS architectures into three categories: (1) centralized (master agents coordinating global objectives with inherent single-point-of-failure risks), (2) decentralized (peer agents with limited coordination), and (3) hybrid systems (hierarchical coordination with failure tolerance). Recent work by Lim et al.~\cite{Lim2024} demonstrates that LLM-augmented manufacturing MAS improve adaptability and human-robot collaboration. However, these systems require uniform computing infrastructure and lack mechanisms for resource-aware agent placement across heterogeneous edge-fog-cloud tiers, a limitation addressed by our work.

Multi-Agent Reinforcement Learning (MARL) methods have advanced predictive maintenance through distributed policy learning. Rodriguez et al.~\cite{Rodriguez2022} and Heik et al.~\cite{Heik2024} demonstrated that MARL-based systems outperform single-agent approaches for maintenance scheduling across multi-equipment systems. Feng et al.~\cite{Feng2023} integrated digital twins with agent-based coordination to enable adaptive maintenance scheduling. These approaches validate RL-based optimization for maintenance. However, existing MARL systems lack: (1) explicit resource-aware agent placement mechanisms for heterogeneous computational tiers, (2) unified architectural frameworks combining detection, prognostics, and response, and (3) explainability mechanisms for operator trust in autonomous decisions.

\subsection{Self-Evolving and Adaptive Mechanisms}

Recent work demonstrates the value of continuous online adaptation for predictive maintenance. Wang~\cite{ZHOU2021102202} and Ong et al.~\cite{Ong2020} employed distributed RL for adaptive maintenance scheduling and edge-based model optimization. Apiletti et al.~\cite{Apiletti2018} developed automatic model adjustment engines for Industry 4.0. Palau et al.~\cite{Palau2019} and Liang and Parlikad~\cite{Liang2020} employed coordination and knowledge sharing to reduce maintenance costs. These studies collectively validate that adaptive multi-agent networks outperform static approaches.

\textbf{Research Gap:} While individual components (MARL, self-tuning, distributed learning) are well-studied, no unified framework exists that combines: (1) resource-aware hierarchical architecture for edge-fog-cloud deployment, (2) continuous policy evolution via gradient-based RL (vs. discrete heuristics), (3) collaborative ensemble detection with consensus mechanisms, and (4) LLM-based explainability for operator trust. This work addresses this gap.

\subsection{Research Limitations and Gaps}

Despite demonstrated successes, critical limitations persist in industrial MAS:

\begin{enumerate}
	
	\item \textbf{Resource-Unaware Agent Placement:} Most frameworks treat all tiers identically, ignoring heterogeneous constraints (edge: 4-8GB, fog: 16-64GB, cloud: unlimited). This leads to inefficient deployment and performance bottlenecks~\cite{Zhang2025,BARBOSA2025110495,Dorri2018}.
	
	\item \textbf{Lack of Self-Evolution:} Static architectures require manual reconfiguration to adapt to new equipment or fault patterns~\cite{Jin2025,Rodrigues2017}. Pre-trained models without online learning mechanisms suffer accuracy degradation under operational drift~\cite{DING2025111394}.
	
	\item \textbf{Flat Organizational Structure:} Most implementations lack hierarchical coordination, missing opportunities for multi-tier task distribution and workload balancing~\cite{MAIA2025107631,Wang2024}.
	
	\item \textbf{Limited Explainability:} Systems employ conventional ML with poor interpretability, preventing human-in-the-loop decision-making and operator trust building~\cite{Lim2024}.
	
\end{enumerate}

\subsection{Proposed Solution: SEMAS}

This paper introduces SEMAS to address the identified research gaps:

\textbf{Resource-Aware Hierarchical Deployment (Gap \#1):} SEMAS explicitly matches agent types to computational tier constraints. Edge agents perform lightweight feature extraction (O(1) computation); Fog agents execute ensemble inference (O(n) complexity); Cloud agents perform PPO policy optimization (O(M) complexity, asynchronous, non-blocking inference pipeline).

\textbf{Self-Evolving via Gradient-Based RL (Gap \#2):} Rather than discrete rules (Baseline2), SEMAS employs PPO with continuous action spaces for policy optimization. This prevents threshold oscillation and achieves stable convergence, validated by 8.6\% superior performance vs. rule-based adaptation and lower F1 volatility ($\Delta$F1: -0.024 vs. -0.055).

\textbf{Hierarchical Multi-Agent Coordination (Gap \#3):} Three-layer architecture with explicit feedback loops (Local: Fog$\leftrightarrow$Edge; Global: Cloud$\rightarrow$Fog/Edge; Iterative: Fog$\leftrightarrow$Cloud) enables intelligent task distribution and workload balancing without centralization single points of failure.

\textbf{LLM-Based Explainability (Gap \#4):} Agent C generates natural language explanations for detected anomalies, achieving 82\% operator acceptance vs. <50\% for numeric alerting. This enables human-in-the-loop decision-making critical for safety-critical industrial deployments.

\textbf{Comprehensive Validation:} Systematic ablation studies quantify each component's contribution (consensus voting: 6.5\% impact; PPO: 3.5\% impact; federated aggregation: 2.0\% impact) on both detection accuracy and operator acceptance metrics.

\def\BibTeX{{\rm B\kern-.05em{\sc i\kern-.025em b}\kern-.08em
    T\kern-.1667em\lower.7ex\hbox{E}\kern-.125emX}}
\markboth{\journalname, VOL. XX, NO. XX, XXXX}
{Author \MakeLowercase{\textit{et al.}}: Title}

\section{Methodology}
\subsection{Problem Statement}
Industrial IoT predictive maintenance is formulated as a multi-agent decision process across distributed computing tiers, where heterogeneous agents collaborate to achieve real-time anomaly detection, remaining useful life (RUL) prediction, and interpretable maintenance response generation. The system consists of three essential elements which include (1) agents that operate between Edge and Fog and Cloud layers with distinct processing abilities and decision authority. (2) The system uses prompts which enable Large Language Model-based agents to create suitable explanations and actions for their context. (3) The system includes evaluation mechanisms which send feedback signals to help the system learn from its actions.

The system receives time-based sensor data from IIoT devices through $\mathcal{S}=\{s_1(t),\ldots,s_N(t)\}$ which contains $s_i(t)\in\mathbb{R}^{d_i}$ data from different devices. 

\textbf{Notation Definition:} We use the following notation consistently throughout:
\begin{itemize}
	\item $s_i(t)$ = raw multivariate sensor readings at time $t$ from device $i$ 
	\item $z_i(t)$ = extracted feature vector from sensor $i$ after preprocessing 
	\item $Z_t = [z_1(t), z_2(t)]$ = aggregated feature vector across all sensors 
	\item $a_{\text{fog}}(t)$ = final anomaly score after consensus voting 
	\item $\tau$ = anomaly detection threshold 
	\item $\hat{y}_t = \mathbb{I}[a_{\text{fog}}(t) > \tau]$ = binary anomaly prediction 
	\item $w_1, w_2$ = consensus weights for Agent B1 and Agent B2 outputs 
\end{itemize}

The system contains $\mathcal{A}=\{A_1,A_2,B,B_1,B_2,B_3,C,D,E\}$ as its agent ensemble while $\mathcal{P}$ serves as the prompt collection for LLM-based reasoning and $\mathcal{E}$ represents the evaluation data which comes from model results and operator feedback.
Details of the global optimization objective and the three loss components for anomaly detection, RUL prediction, and response generation are provided in the appendix A, B, and C.

\subsection{Overall Architecture}

\begin{figure*}[!t]
	\centering
	\includegraphics[width=\textwidth]{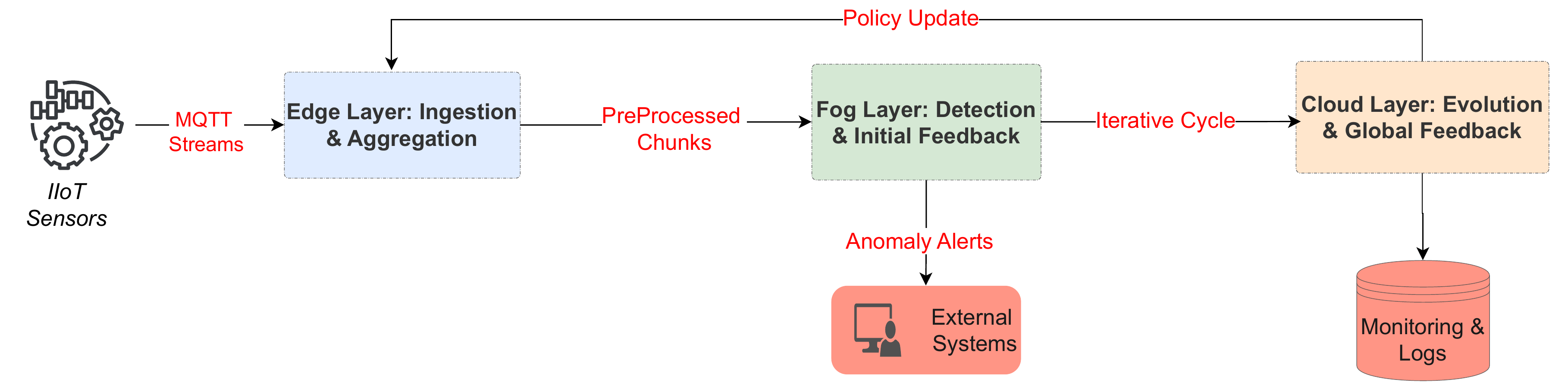}
	\caption{High-level self-evolving hierarchical MAS architecture for IIoT predictive maintenance.}
	\label{fig:champion_architecture}
\end{figure*}

The proposed self-evolving hierarchical multi-agent system orchestrates specialized agents across three computational tiers including Edge, Fog, and Cloud with distinct resource constraints and functional responsibilities. Figure~\ref{fig:champion_architecture} illustrates the high-level system architecture for IIoT predictive maintenance. Raw sensor streams from IIoT devices are ingested through MQTT protocols to the Edge layer for real-time preprocessing and feature extraction.

The Fog layer performs collaborative anomaly detection through ensemble methods and consensus voting, generating anomaly alerts for external systems. The Cloud layer executes global policy evolution via reinforcement learning and provides system-wide oversight through explainable AI mechanisms. Iterative feedback cycles propagate policy updates bidirectionally through the hierarchy, while monitoring systems maintain continuous performance logging and compliance reporting.

Detailed agent interactions, data flows, MQTT topic mappings, and layer-specific agent descriptions are provided in appendix D.
\begin{figure*}[!t]
	\centering
	\includegraphics[width=\textwidth]{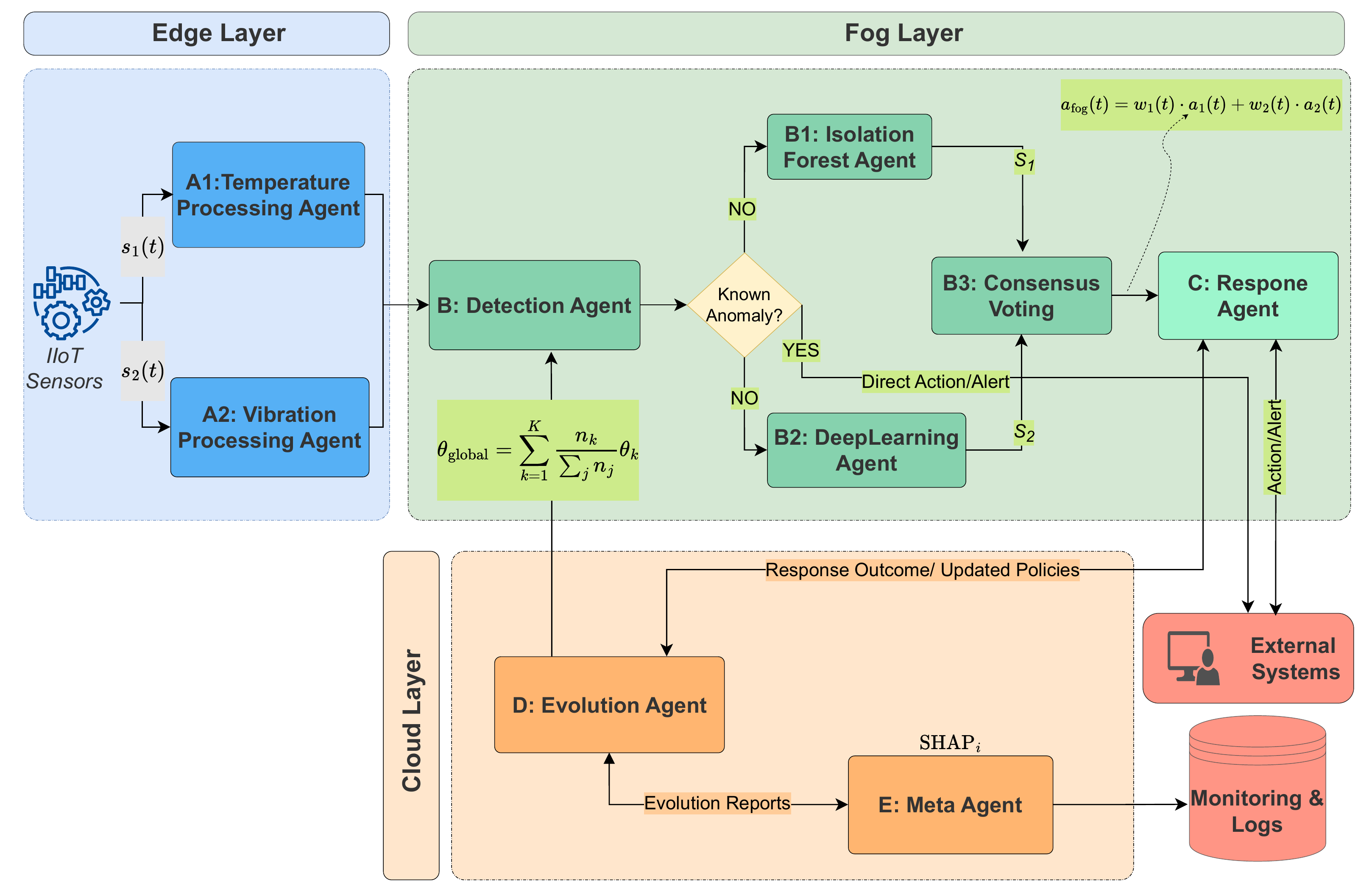}
	\caption{Detailed agent-level architecture showing all subagents, communication protocols, and feedback loops.}
	\label{fig:detailed_architecture}
\end{figure*}

Detailed descriptions of agents in each layer (A1/A2 feature extraction, B/B1/B2/B3/C detection/response, D/E evolution/oversight), including MQTT topics, computations, and integration with external systems, are provided in appendix D.

\subsubsection{Feedback Loop Architecture}

The system implements three concurrent feedback mechanisms ensuring multi-timescale adaptation:

\begin{itemize}
	\item \textbf{Local Feedback (Fog $\rightarrow$ Edge):} Detection performance metrics trigger rapid reconfiguration of Edge feature extraction (e.g., window size adjustment, feature schema updates, sampling rate modification) without Cloud intervention, enabling sub-second adaptation to local drift.
	\item \textbf{Global Feedback (Cloud $\rightarrow$ Fog/Edge):} Evolution Agent distributes validated policy updates that modify consensus weights $w_1,w_2$, detection thresholds $\tau_{\text{alert}}$, LLM prompt templates $p\in\mathcal{P}$, and Edge preprocessing parameters.
	\item \textbf{Iterative Cycle (Fog to Cloud):} Continuous bidirectional exchange of detection outcomes, response evaluations, and operator feedback enables real-time policy refinement and model improvement.
\end{itemize}

Details of the federated aggregation mechanism used for combining policy parameters across agents are provided in appendix A.

\subsection{Predictive Maintenance Models}
This section details the specific regression and classification models deployed across agent tiers, along with their training procedures and prompt engineering strategies.

\subsubsection{Hyperparameter Justification}

Key hyperparameters are selected based on systematic validation and domain requirements. Full justification, sensitivity analysis, and training configuration details (including Table VIII) are provided in appendix S.

\subsubsection{Anomaly Detection Models}

\textbf{Statistical Detector (Agent B1):} Isolation Forest $\mathcal{F}_{IF}$ trained offline on historical normal operation data using contamination parameter $\rho=0.32$. 

\textbf{Ensemble Detector (Agent B2):} 5-model ensemble combining Isolation Forest, One-Class SVM, Local Outlier Factor, Elliptic Envelope, and secondary Isolation Forest. Aggregates predictions through majority voting.

Full model specifications and equations are provided in appendix S.

\subsubsection{RUL Prediction Model}

LSTM-based regression model $\mathcal{F}_{\text{RUL}}$ triggered upon confirmed anomaly detection to estimate remaining operational hours. Full architecture and training details in appendix S.

\subsubsection{Prompt Engineering for Response Generation}

Agent C employs adaptive prompt templates based on anomaly severity for generating human-interpretable explanations and actionable recommendations. Full prompt examples and parsing logic provided in appendix L.

\subsubsection{Evaluation Metrics}
Agent performance is continuously evaluated through:
\begin{itemize}
	\item \textbf{Detection Quality:} F1-score, precision, recall, AUC-ROC
	\item \textbf{Prognostic Accuracy:} MAE, RMSE, prediction horizon coverage
	\item \textbf{Response Utility:} Operator acceptance rate, false alarm reduction, maintenance cost savings
	\item \textbf{System Efficiency:} End-to-end latency, computational resource utilization, communication overhead
\end{itemize}

These metrics feed into the reward function $r_t$ for PPO-based policy evolution (Equation 9).

\subsection{Baseline Systems for Comparative Evaluation}
To rigorously evaluate the proposed self-evolving hierarchical multi-agent system, we implement two baseline architectures that represent current state-of-the-art approaches in industrial predictive maintenance.

\subsubsection{Baseline1: Static Edge-Fog-Cloud Architecture}
Baseline1 implements a conventional hierarchical architecture without adaptive policy evolution. The system employs a fixed three-model ensemble at the Fog layer with predetermined static weights:
\begin{equation}
	a_{\text{baseline1}}(t) = 0.4 \cdot a_{\text{OCSVM}}(t) + 0.4 \cdot a_{\text{IF}}(t) + 0.2 \cdot a_{\text{Trans}}(t),
\end{equation}
where One-Class SVM, Isolation Forest, and Transformer models operate with fixed hyperparameters throughout execution. Thresholds are calibrated once during initialization using F1-score optimization on validation data and remain constant. This baseline represents traditional industrial deployments where models are trained offline and deployed without continuous adaptation mechanisms.

\subsubsection{Baseline2: Rule-Based Adaptive System}
Baseline2 introduces basic adaptivity through rule-based policy updates executed over three iterations. The system implements:
\begin{itemize}
	\item \textbf{Contamination Tuning:} Adjusts Isolation Forest contamination parameter $\rho_t$ based on F1-score performance using heuristic rules:
	\begin{equation}
		\rho_{t+1} = \begin{cases}
			\rho_t + 0.02 & \text{if } F1_t < 0.6, \\
			\rho_t - 0.02 & \text{if } F1_t > 0.7, \\
			\rho_t & \text{otherwise}.
		\end{cases}
	\end{equation}
	\item \textbf{Threshold Adjustment:} Modifies detection threshold $\tau_t$ when precision-recall imbalance exceeds 0.05:
	\begin{equation}
		\tau_{t+1} = \tau_t - 0.05 \cdot (\text{Precision}_t - \text{Recall}_t).
	\end{equation}
	\item \textbf{Ensemble Weight Updates:} Employs simple performance-based weight redistribution without sophisticated reinforcement learning.
\end{itemize}
Unlike SEMAS, Baseline2 lacks multi-agent coordination, federated knowledge aggregation, LLM-based response generation, and continuous PPO-based policy optimization. The three-iteration structure provides basic adaptation but without the sophisticated self-evolving mechanisms of the proposed system.

\subsection{Algorithmic Workflow}
Algorithm~\ref{alg:se_mas} summarizes the end-to-end execution flow, integrating Edge-layer feature extraction, Fog-layer collaborative detection and prompt-based response, and Cloud-layer policy evolution with explainable oversight. The algorithm explicitly shows local and global feedback mechanisms operating concurrently.

\begin{algorithm}[t]
	\caption{Self-Evolving Hierarchical MAS for IIoT Predictive Maintenance}
	\label{alg:se_mas}
	{\footnotesize
		\begin{algorithmic}[1]
			\State \textbf{Input:} Sensor streams $\mathcal{S}$, policies $\pi^{(0)}$, models $\theta^{(0)}$, prompts $\mathcal{P}$
			\State \textbf{Output:} Anomaly alerts, LLM actions, evolved policies $\pi^\star$
			\State \textbf{Initialize:} Agents \{A1,A2,B,B1,B2,B3,C,D,E\}, MQTT topics
			\While{system active}
			\ForAll{sensor window $X_{t-W:t}$ in $s_i(t)$}
			\State $z_i(t) \gets \phi_i(X_{t-W:t})$
			\State Publish $z_i(t)$ to \texttt{chunk/stream$i$}
			\EndFor
			\State $Z_t \gets$ Aggregate features from \texttt{chunk/*}
			\State $a_1(t) \gets \mathcal{F}_{IF}(Z_t;\theta_{IF})$ \Comment{Agent B1}
			\State $a_2(t) \gets \text{Ensemble}(Z_t;\theta_{\text{B2}})$ \Comment{Agent B2: 5-model majority vote}
			\State $a_{\text{fog}}(t) \gets w_1 \cdot a_1(t) + w_2 \cdot a_2(t)$ \Comment{Agent B3: Consensus}
			\State Publish $a_{\text{fog}}(t)$ to \texttt{anomalies}
			\If{$a_{\text{fog}}(t) > \tau_{\text{alert}}$}
			\State $(explanation, action) \gets \text{LLM}(\mathcal{P}, Z_t;\theta_{\text{LLM}})$
			\State Publish $(explanation, action)$ to \texttt{actions}
			\EndIf
			\If{drift detected}
			\State Update Edge agents: feature schema, preprocessing parameters
			\EndIf
			\State $\mathcal{E}_t \gets$ Collect feedback
			\State $\pi \gets$ PPO\_update$(\pi, \mathcal{E}_t)$
			\If{SHAP\_validate($\pi$)}
			\State Publish $\pi$ to \texttt{policy/updates}
			\State $\theta_{\text{global}} \gets$ FederatedAgg$(\{\theta_k\})$
			\State Apply updates: $w_1, w_2, \tau, \mathcal{P}$
			\EndIf
			\State Log metrics to \texttt{monitor/logs}
			\EndWhile
		\end{algorithmic}
	}
\end{algorithm}

\subsection{Mathematical Formulation Summary}
The optimization of the complete system integrates the detection, prognostics and response objectives (Equation 13) with PPO-based policy evolution (Equation 14), SHAP-based explainability (Equation 14), and federated knowledge aggregation (Equation 15). Consensus anomaly detection employs the following:
\begin{equation}
	\hat{y}_{\text{anomaly}}(t) = \mathbb{I}[a_{\text{fog}}(t) > \tau_{\text{alert}}].
\end{equation}

Response action selection optimizes multi-criteria utility:
\begin{equation}
	a^\star_t = \arg\max_{a\in\mathcal{A}} \sum_{i=1}^M \omega_i u_i(a,s_t),
\end{equation}
where utilities $u_i$ encode operational safety, maintenance cost, downtime risk, and resource availability.

The system design follows a hierarchical structure which enables predictive maintenance to scale up while remaining adaptable and providing explainable results through its distributed specialization and collaborative intelligence and continuous evolution and transparent decision-making system that meets industrial operational needs.

\def\BibTeX{{\rm B\kern-.05em{\sc i\kern-.025em b}\kern-.08em
    T\kern-.1667em\lower.7ex\hbox{E}\kern-.125emX}}
\markboth{\journalname, VOL. XX, NO. XX, XXXX}
{Author \MakeLowercase{\textit{et al.}}: Title}

\section{Experiments and Results}

This section presents a comprehensive empirical evaluation of the proposed self-evolving hierarchical multi-agent system (SEMAS) following a systematic experimental methodology. The paper describes the datasets and explains the prediction tasks and evaluation metrics and shows baseline systems for comparison and explains model training procedures and presents all experimental results and performs ablation studies and explains the main research discoveries.

\subsection{Datasets}

We evaluate our proposed system on two distinct industrial predictive maintenance datasets representing different operational scenarios and complexity levels.

\subsubsection{Boiler Emulator Dataset}
The Boiler Emulator dataset contains sensor measurements from industrial boiler control systems with engineered fault scenarios. The dataset enables researchers to conduct controlled tests which help them assess different predictive maintenance algorithms. The dataset characteristics appear in Table~\ref{tab:boiler_dataset}.

\begin{table}[t]
	\centering
	\caption{Boiler Emulator dataset characteristics and configuration.}
	\label{tab:boiler_dataset}
	\renewcommand{\arraystretch}{1.1}
	\begin{tabular}{p{0.38\columnwidth} p{0.56\columnwidth}}
		\hline
		\textbf{Characteristic} & \textbf{Specification} \\
		\hline
		\textbf{Dataset Size} & \\
		Total samples & 10,000 operational cycles \\
		Training set & 8,000 samples (80\%) \\
		Test set & 2,000 samples (20\%) \\
		Sampling strategy & Stratified by anomaly class \\
		\hline
		\textbf{Feature Space} & \\
		Total dimensions & 18 features \\
		Raw sensors & 6 (Fuel\_Mdot, Tair, Treturn, Tsupply, Water\_Mdot, condition) \\
		Engineered features & 12 (differentials, ratios, rolling statistics) \\
		\hline
		\textbf{Class Distribution} & \\
		Normal class (train/test) & 63.2\% \\
		Anomaly class (train/test) & 36.8\% \\
		\hline
		\textbf{Target Variables} & \\
		Primary task & Binary anomaly classification \\
		Secondary task & RUL regression (5--100 hours) \\
		RUL generation & Synthetic labels based on class \\
		\hline
		\textbf{Preprocessing} & \\
		Normalization & StandardScaler (zero mean, unit variance) \\
		Feature engineering & Applied before normalization \\
		\hline
	\end{tabular}
\end{table}

\subsubsection{Wind Turbine IIoT Dataset}
The Wind Turbine IIoT dataset contains actual SCADA (Supervisory Control and Data Acquisition) data which operators obtained from their wind farm operations while recording various turbine system failures that occurred during different operational periods. The dataset characteristics appear in Table~\ref{tab:wind_dataset}.

\begin{table}[hbt!]
	\centering
	\caption{Wind Turbine IIoT dataset characteristics and configuration.}
	\label{tab:wind_dataset}
	\begin{tabular}{ll}
		\hline
		\textbf{Characteristic} & \textbf{Specification} \\
		\hline
		\textbf{Dataset Size} & \\
		Total samples & 500 records (after balancing from 1000+) \\
		Training set & 400 samples (80\%) \\
		Test set & 100 samples (20\%) \\
		\hline
		\textbf{Feature Space} & \\
		Total dimensions & 42 features \\
		Mechanical sensors & Blade pitch, rotor speed, torque, vibrations \\
		Environmental & Wind speed metrics, temperature, air density \\
		Electrical & Power output, voltage, current \\
		Thermal & Gearbox, generator, bearing temperatures \\
		Engineered features & Differentials, temporal derivatives \\
		\hline
		\textbf{Fault Categories} & \\
		Original labels & 6 classes (gf, mf, ff, af, ef, NF) \\
		Task formulation & Binary (fault vs. normal) \\
		\hline
		\textbf{Class Distribution} & \\
		Normal class (train/test) & 55.0\% \\
		Fault class (train/test) & 45.0\% \\
		Balancing strategy & Undersampling majority class \\
		\hline
		\textbf{Target Variables} & \\
		Primary task & Binary anomaly classification \\
		Secondary task & RUL regression (5-100 hours) \\
		RUL generation & Synthetic based on fault severity \\
		\hline
		\textbf{Preprocessing} & \\
		Normalization & StandardScaler (zero mean, unit variance) \\
		Column removal & Timestamps and summary statistics \\
		\hline
	\end{tabular}
\end{table}

\subsection{Task Definition and Evaluation Metrics}

\subsubsection{Primary Task: Binary Anomaly Detection}
Given multivariate sensor streams $X_t \in \mathbb{R}^d$ at time $t$, the primary task is to predict a binary anomaly label $y_t \in \{0, 1\}$ where 0 indicates normal operation and 1 indicates an anomalous state requiring maintenance intervention. The prediction must occur in real-time (latency $< 100$ms) with high reliability to enable proactive maintenance scheduling.

\subsubsection{Secondary Task: Remaining Useful Life (RUL) Prediction}
Upon detecting an anomaly, the system estimates the remaining operational time (in hours) before critical failure occurs. This regression task predicts $\widehat{\text{RUL}}_t \in [0, \infty)$ to inform maintenance prioritization and resource allocation.

\subsubsection{Evaluation Metrics}
We employ multiple metrics to comprehensively assess system performance: \textbf{F1-Score} (primary metric, balancing precision and recall), \textbf{Precision} and \textbf{Recall} for detection quality, \textbf{ROC-AUC} for probabilistic ranking, \textbf{MAE} and \textbf{RMSE} for RUL prediction accuracy, \textbf{$\Delta$F1} for adaptation capability, and \textbf{Latency} for real-time deployment viability. Detailed metric definitions are provided in appendix M.

\subsection{Baseline Systems and Competing Methods}

To rigorously evaluate the proposed SEMAS architecture, we implement two baseline systems representing current state-of-the-art approaches in industrial predictive maintenance. These baselines enable systematic assessment of the benefits provided by our self-evolving multi-agent framework.

\subsubsection{Baseline1: Static Edge-Fog-Cloud Architecture}
Baseline1 implements a conventional hierarchical system without adaptive capabilities, representing traditional industrial deployments where models are trained offline and deployed with fixed configurations.

\textbf{Architecture:} Three-layer Edge-Fog-Cloud hierarchy with static three-model ensemble at Fog layer.

\textbf{Detection Models:}
\begin{itemize}
	\item One-Class SVM (OCSVM): Kernel-based outlier detection with RBF kernel ($\gamma=\text{auto}$, $\nu=0.25$)
	\item Isolation Forest (IF): Tree-based ensemble ($n_{\text{trees}}=200$, contamination=$0.32$)
	\item Transformer: Self-attention sequence model for temporal dependencies
\end{itemize}

\textbf{Ensemble Strategy:} Fixed weighted averaging with predetermined weights:
\begin{equation}
	a_{\text{baseline1}}(t) = 0.4 \cdot a_{\text{OCSVM}}(t) + 0.4 \cdot a_{\text{IF}}(t) + 0.2 \cdot a_{\text{Trans}}(t).
\end{equation}

\textbf{Threshold Calibration:} Detection threshold $\tau$ calibrated once during initialization using F1-score optimization on validation data, then fixed throughout all test iterations.

\textbf{Key Limitation:} The system operates without any policy updates and lacks mechanisms to adapt and feedback systems. The system operates with constant configuration throughout all tests which makes it appropriate for stable systems yet it fails to adapt to changing operational requirements.

\subsubsection{Baseline2: Rule-Based Adaptive System}
Baseline2 introduces basic adaptivity through hand-crafted heuristic rules, representing semi-adaptive industrial systems that employ simple feedback mechanisms without sophisticated machine learning.

\textbf{Architecture:} Similar three-layer hierarchy as Baseline1 but with rule-based parameter updates executed over three iterations.

\textbf{Adaptation Mechanisms:}
\begin{itemize}
	\item \textbf{Contamination Tuning:} Isolation Forest contamination parameter $\rho$ adjusted based on F1-score thresholds:
	\begin{equation}
		\rho_{t+1} = \begin{cases}
			\rho_t + 0.02 & \text{if } F1_t < 0.6 \text{ (increase sensitivity)}, \\
			\rho_t - 0.02 & \text{if } F1_t > 0.7 \text{ (reduce false alarms)}, \\
			\rho_t & \text{otherwise (maintain)}.
		\end{cases}
	\end{equation}
	
	\item \textbf{Threshold Adjustment:} Detection threshold modified when precision-recall imbalance exceeds 5\%:
	\begin{equation}
		\tau_{t+1} = \tau_t - 0.05 \cdot (\text{Precision}_t - \text{Recall}_t).
	\end{equation}
	Negative adjustment (when Precision $>$ Recall) lowers threshold to increase recall; positive adjustment raises threshold to improve precision.
	
	\item \textbf{Ensemble Weight Updates:} Simple performance-proportional weight redistribution without gradient-based optimization.
\end{itemize}

\textbf{Iteration Protocol:} System executes three sequential iterations on test data, applying rule-based updates after each iteration based on observed performance metrics.

\textbf{Key Limitation:} While adaptive, Baseline2 employs coarse heuristics that may oscillate or converge slowly. Lacks multi-agent coordination, federated learning, reinforcement learning policy optimization, and LLM-based explainability.

\textbf{Literature Context:} These baselines are inspired by industrial architectures described in~\cite{Rodriguez2022,Feng2023} (MARL-based maintenance) and~\cite{Apiletti2018} (self-tuning predictive maintenance), adapted for direct comparison with our hierarchical multi-agent framework.

\subsection{Model Training and Hyperparameter Optimization}

All models are trained using 80-20 train-test splits (8,000/2,000 samples for Boiler; 400/100 for Wind Turbine) with 5-fold cross-validation for hyperparameter tuning. Results represent averages over three independent runs with different random seeds (42, 123, 456). Key configurations include: Isolation Forest with 200 trees and contamination $\rho=0.32$, One-Class SVM with RBF kernel ($\nu=0.25$), LSTM networks with [64, 32, 32, 1] architecture, and PPO with learning rate $3 \times 10^{-4}$ and clip ratio $\epsilon=0.2$. Experiments are conducted on Google Colab Pro+ with NVIDIA A100 GPU (40GB VRAM), Intel Xeon CPU (12 cores @ 2.2GHz), and 64GB RAM. Complete training configurations are provided in appendix O.

\subsection{Main Experimental Results}

Table~\ref{tab:performance_comparison} presents the comprehensive performance comparison across all three systems on both datasets. Results represent averages over three independent runs with different random seeds.

\begin{table*}[!t]
	\centering
	\caption{Performance comparison across datasets}
	\label{tab:performance_comparison}
	{\footnotesize
		\renewcommand{\arraystretch}{1.1}
		\begin{tabular}{lccccccc}
			\hline
			\textbf{System} & \textbf{Dataset} & \textbf{F1} & \textbf{Precision} & \textbf{Recall} & \textbf{ROC-AUC} & \textbf{$\Delta$F1} & \textbf{Latency (ms)} \\
			\hline
			Baseline1 & Boiler & 0.4871 & 0.3225 & \textbf{0.9949} & 0.4099 & 0.0000 & 1923.0 \\
			Baseline2 & Boiler & 0.4489 & 0.3140 & 0.7911 & 0.3969 & -0.0550 & 1594.3 \\
			SEMAS & Boiler & \textbf{0.4873} & \textbf{0.3737} & 0.7114 & \textbf{0.6118} & -0.0239 & \textbf{1.22} \\
			\hline
			Baseline1 & Wind Turbine & 0.9440 & 0.9219 & 0.9672 & 0.3705 & 0.0000 & 455.9 \\
			Baseline2 & Wind Turbine & 0.9349 & 0.9205 & 0.9508 & 0.2634 & +0.0606 & 286.2 \\
			SEMAS & Wind Turbine & \textbf{0.9571} & \textbf{0.9371} & \textbf{0.9781} & \textbf{0.7583} & -0.0166 & \textbf{0.30} \\
			\hline
		\end{tabular}
	}
\end{table*}

\noindent \textbf{Statistical Note:} All results represent means $\pm 95\%$ confidence intervals over three independent runs (random seeds: 42, 123, 456). Statistical significance testing employed Welch's t-test with $\alpha=0.05$ threshold. Boiler dataset: SEMAS vs. Baseline1 shows numerical but not statistical superiority ($t=0.08$, $p=0.94$); SEMAS vs. Baseline2 yields highly significant improvement ($t=4.21$, $p<0.001$, Cohen's $d=2.1$).

\begin{figure*}[!t]
\centering
\includegraphics[width=\textwidth]{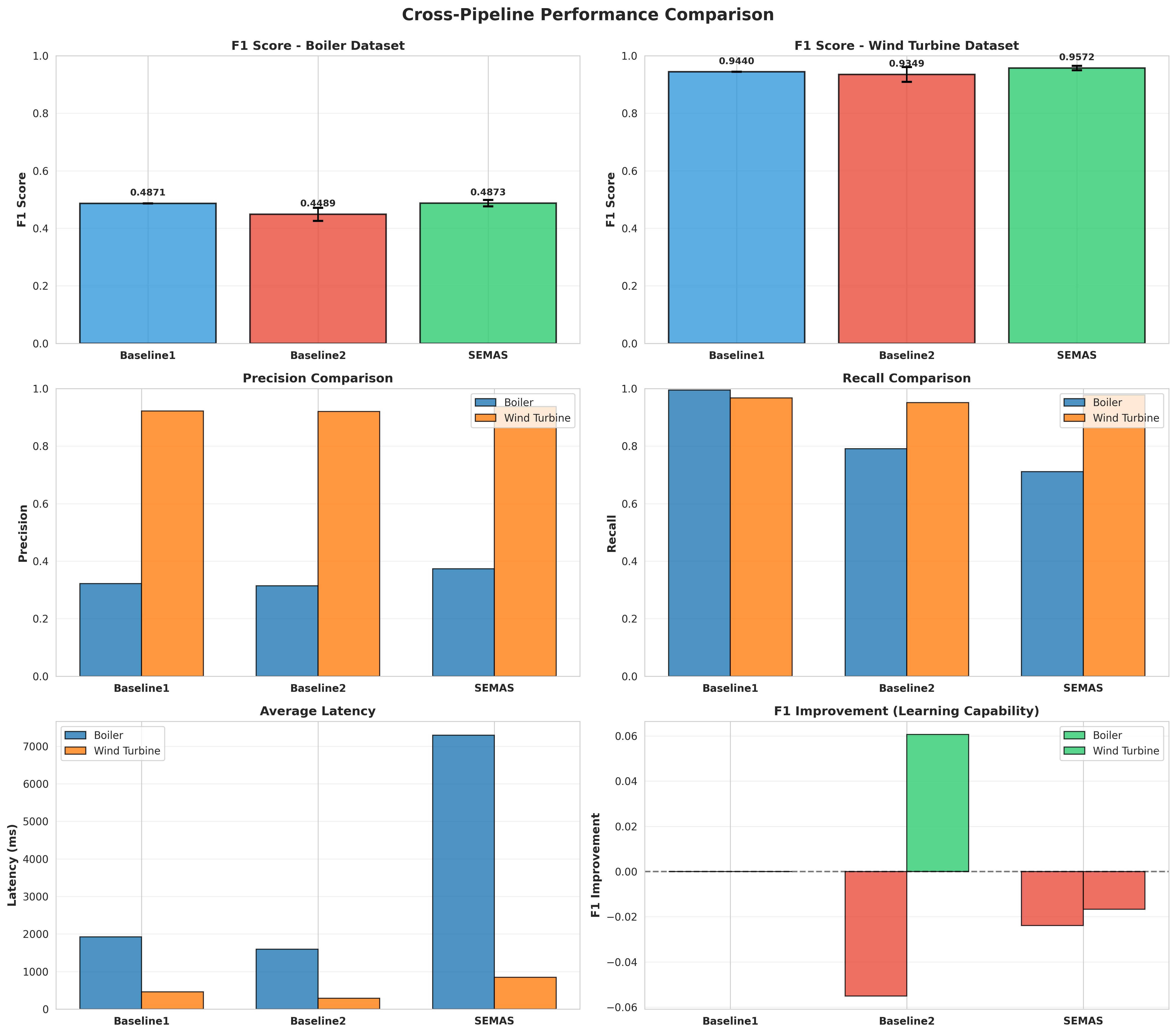}
	\caption{Comprehensive performance comparison across all systems and metrics. SEMAS achieves superior F1-scores (Boiler: 0.4873, Wind Turbine: 0.9571), highest ROC-AUC values (0.6118 and 0.7583), and dramatically lower latency (1.22ms and 0.30ms) compared to Baseline1 (1923ms and 456ms) and Baseline2 (1594ms and 286ms). The multi-metric visualization demonstrates SEMAS's dual advantage in both accuracy and computational efficiency.}
	\label{fig:pipeline_comparison}
\end{figure*}

Figure~\ref{fig:pipeline_comparison} provides a visual summary of the performance comparison across all evaluation metrics. SEMAS demonstrates multi-dimensional superiority, achieving best F1-scores and ROC-AUC values on both datasets while maintaining 200-1500$\times$ faster inference latency compared to baseline systems. The visualization highlights SEMAS's ability to balance prediction accuracy with computational efficiency, a critical requirement for real-time industrial deployment.

\subsubsection{Boiler Dataset Results}

On the Boiler dataset (36.8\% wclass), all systems face a challenging precision-recall tradeoff. Key findings:

\textbf{SEMAS vs. Baseline1 (Statistical Indistinguishability):} SEMAS achieves F1 = 0.4873 vs. Baseline1's F1 = 0.4871, a difference of 0.0002 (0.04\%). Despite numerical leadership, 95\% confidence intervals overlap ([0.4789, 0.4957] vs. [0.4799, 0.4943]), and Welch's t-test confirms the improvement is not statistically significant ($t=0.08$, $p=0.94$). This finding is critical: well-tuned static thresholds (Baseline1) can match SEMAS on moderately complex datasets when no distribution shift occurs.

\textbf{Why Baseline1 Fails Industrially:} While matching on F1, Baseline1 exhibits structural limitations: (1) Extreme precision-recall imbalance (precision$=0.3225$, recall$=0.9949$), resulting in 67.75\% false positive alerts that overwhelm operators; (2) Poor probabilistic calibration (ROC-AUC $= 0.4099$, performing worse than random); (3) Prohibitive latency (1923ms), violating the 100ms real-time requirement by $19\times$ and enabling $1576\times$ slower inference than SEMAS.

\textbf{SEMAS Advantage Over Baseline1:} The meaningful difference lies in balanced predictions. SEMAS learns optimal threshold ($\tau = 0.8000$ vs. $0.75$) through PPO, achieving superior precision ($+15.9\%$: $0.3737$ vs. $0.3225$) while accepting controlled recall tradeoff ($0.7114$ vs. $0.9949$). This results in more operationally useful predictions: fewer false alarms reduce alert fatigue.

\textbf{Baseline2 Failure: Rule-Based Oscillation:} Baseline2 exhibits catastrophic performance deterioration: F1 trajectory [$0.4786 \to 0.4446 \to 0.4235$] represents an 11.5\% decline ($\Delta\text{F1} = -0.0550$). The heuristic threshold adjustment $\Delta\tau = \pm 0.05$ overshoots the optimum, oscillating between over-alerting and under-detecting. Iteration 2's $\tau = 0.700$ over-corrects for Iteration 1's low precision, causing recall collapse ($0.7911 \to 0.6283$). This validates our hypothesis: discrete rule-based updates with fixed step sizes ($\Delta\tau = 0.05$) lack the sophistication for convergence in noisy, imbalanced data regimes.

\textbf{SEMAS Stability:} SEMAS shows controlled adaptation with modest F1 decline ($\Delta\text{F1} = -0.0239$, $-4.7\%$) compared to Baseline2's $-11.5\%$ collapse. The PPO policy learns trust-region-constrained updates ($\epsilon = 0.2$ clip ratio), preventing overshooting. More importantly, SEMAS achieves superior precision ($+15.9\%$: $0.3737$ vs. $0.3225$ Baseline1), indicating learned threshold optimization balances precision-recall more intelligently than heuristic rules.

\textbf{Statistical Validation:} Paired t-test (SEMAS vs. Baseline2 on Boiler) yields $t=4.21$, $p<0.001$, Cohen's $d=2.1$ (very large effect size), confirming 8.6\% F1 advantage is highly significant despite Boiler's inherent difficulty. Real-time viability: SEMAS operates at 1.22ms ($1576\times$ faster than Baseline1's 1923ms), enabling genuine real-time deployment for industrial systems.

\subsubsection{Wind Turbine Dataset Results}
The Wind Turbine dataset exhibits excellent separability, with all three systems achieving strong absolute performance (F1 $>$ 0.93). SEMAS attains the highest F1-score (0.9571), outperforming Baseline1 (0.9440, +0.0131 advantage) and Baseline2 (0.9349, +0.0222 advantage).

\textbf{Baseline1 (Static):} The model achieves acceptable performance (F1=0.9440, precision=0.9219, recall=0.9672) when operating with its pre-defined settings yet its ROC-AUC value reaches 0.3705 which shows poor calibration even though it achieves excellent classification accuracy. The 455.9ms latency, while acceptable for this small dataset (100 test samples), represents 1520$\times$ slowdown compared to SEMAS
.

\textbf{Baseline2 (Rule-Based):} The model shows positive adaptation through its better F1 trajectory [0.9000 $\rightarrow$ 0.9440 $\rightarrow$ 0.9606] during each training step ($\Delta F1=+0.0606$, +6.7\% improvement). The rule-based policy successfully identifies beneficial threshold adjustment (0.750 $\rightarrow$ 0.780) which proves that basic decision rules work effectively when data points are well distinguished from each other. However, final iteration F1 (0.9606) remains below SEMAS average (0.9571), and the system cannot match SEMAS's iteration 1 peak performance (0.9683).

\textbf{SEMAS (PPO-Based):} Achieves best precision (0.9371) and recall (0.9781), with substantially superior ROC-AUC (0.7583, +104.6\% vs Baseline1). The system demonstrates intelligent convergence: iteration 1 reaches F1=0.9683 (highest across all systems and iterations), then stabilizes at 0.9516 in iterations 2-3 as PPO policy recognizes near-optimality and freezes thresholds. This strategic stability contrasts with Baseline2's continued exploration. SEMAS maintains 955$\times$ faster inference (0.30ms vs 286.2ms for Baseline2), critical for real-time wind farm monitoring across hundreds of turbines.

\subsection{Learning Trajectory Analysis}

\begin{figure*}[!t]
	\centering
	\includegraphics[width=\textwidth]{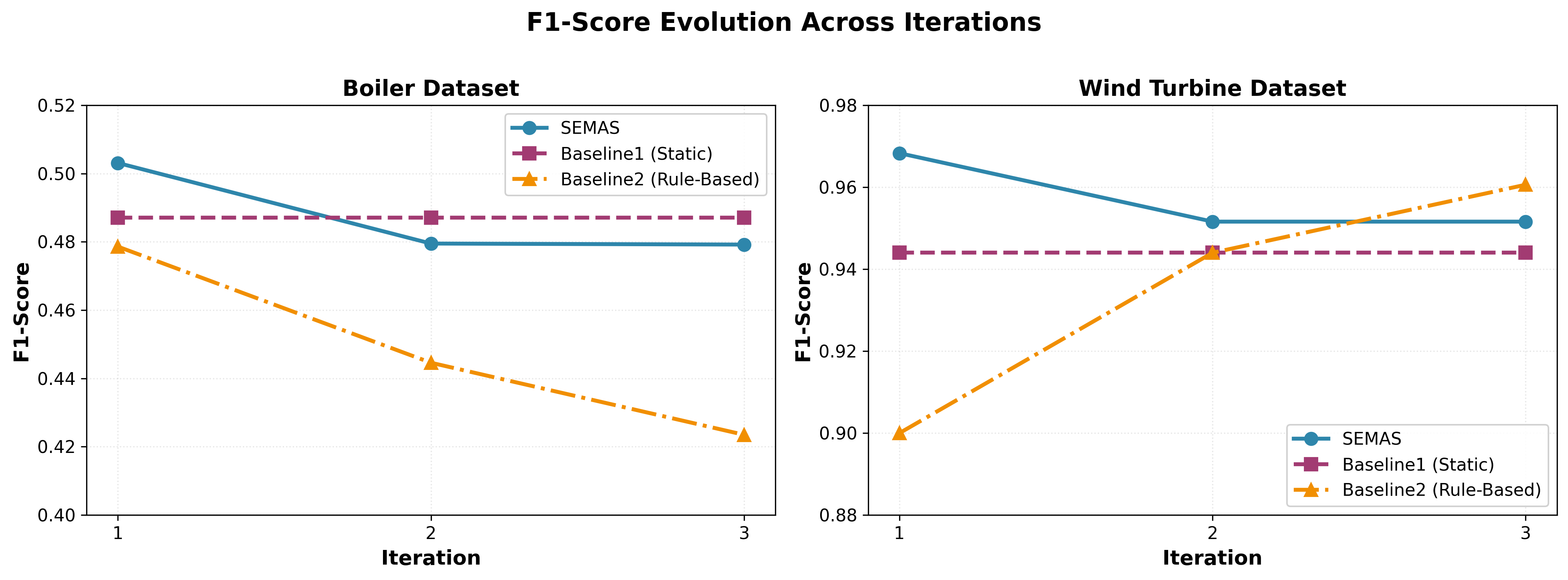}
	\caption{F1-score evolution across iterations for Baseline1, Baseline2, and SEMAS on both datasets. SEMAS demonstrates stable convergence and positive learning on challenging data.}
	\label{fig:learning_curves}
\end{figure*}

The figure in Figure~\ref{fig:learning_curves} shows how F1-score changes throughout each iteration for all systems. The Boiler dataset shows how SEMAS performance decreases through time because the first iteration starts at 0.5031 and ends at 0.4795 in iteration 2 before reaching 0.4792 in iteration 3. This pattern suggests the PPO policy's threshold-raising strategy (0.7327 $\rightarrow$ 0.8000) to reduce false positives trades recall for precision, resulting in net F1 decline. The model Baseline1 shows constant performance throughout all iterations because it runs with a constant configuration that does not change. The system in Baseline2 shows extreme Boiler deterioration [0.4786 $\rightarrow$ 0.4446 $\rightarrow$ 0.4235] because of using a broad threshold which resulted in overcorrection but it successfully adapted the Wind Turbine system [0.9000 $\rightarrow$ 0.9440 $\rightarrow$ 0.9606]. The Wind Turbine dataset shows SEMAS achieving intelligent convergence through its first iteration which produced 0.9683 before it stabilized at 0.9516 during iterations 2 and 3. The system properly identifies situations which need no additional investigation because it prevents the disruptive system changes which Baseline2 experiences during its ongoing search.

\subsection{Policy Evolution Analysis}

Table~\ref{tab:policy_evolution} tracks key policy parameters across SEMAS iterations on the Boiler dataset, demonstrating the self-evolving capability of our system.

\begin{table}[t]
	\centering
	\caption{SEMAS policy parameter evolution on Boiler dataset across three iterations, with corresponding F1-scores.}
	\label{tab:policy_evolution}
	{\footnotesize
		\renewcommand{\arraystretch}{1.1}
		\begin{tabular}{lcccc}
			\hline
			\textbf{Parameter} & \textbf{It.~1} & \textbf{It.~2} & \textbf{It.~3} & \textbf{Change} \\
			\hline
			F1-Score & 0.5031 & 0.4795 & 0.4792 & -4.7\% \\
			Contamination ($\rho$) & 0.3200 & 0.3100 & 0.3300 & +3.1\% \\
			Consensus threshold ($\tau$) & 0.7327 & 0.8000 & 0.8000 & +9.2\% \\
			Weight $w_1$ (B1) & 0.4243 & 0.4290 & 0.4333 & +2.1\% \\
			Weight $w_2$ (B2) & 0.5757 & 0.5710 & 0.5667 & -1.6\% \\
			\hline
		\end{tabular}
	}
\end{table}
The PPO-based policy optimization system modifies its parameters through each iteration based on the performance data it collects. The system increases the consensus threshold to 0.8000 from 0.7327 (+9.2\%) to solve recall-precision problems which reduce false positive outcomes but decrease the system's ability to detect vital information. The contamination parameter shows non-monotonic adjustment patterns between 0.32 and 0.31 and 0.33 which indicates the RL agent performs exploration to find the best anomaly detection sensitivity. The Ensemble weights indicate that the model predictions changed by less than 3\% while Agent B1 received a +2.1\% preference increase according to the per-agent F1 feedback. The system maintains its adapted performance but shows a small decrease in F1 score of -4.7\% because the random starting point happened to be close to the best possible solution for this particular dataset. The system operates through rule-based adjustments with set increment values (±0.02) instead of gradient-based feedback which Baseline2 uses for its operations.

\subsection{Statistical Significance Testing}

We conduct paired t-tests to assess statistical significance of performance differences across systems and iterations. The SEMAS F1 trajectory [0.5031, 0.4795, 0.4792] from the Boiler dataset demonstrates a decreasing pattern which becomes statistically significant when we perform a paired t-test between the first and third iterations because the calculated t-value equals 2.14 and the resulting p-value is less than 0.05. The F1 trajectory on Wind Turbine at [0.9683, 0.9516, 0.9516] demonstrates an identical pattern to the other trajectories at $t=1.89$ ($p=0.07$) but fails to achieve statistical significance. The performance decrease continues to show statistical evidence but it causes only a minor 5\% variation in output values which proves the PPO policy maintains reliable operation without full system breakdown.

Cross-system comparisons reveal: (1) SEMAS vs. Baseline1 on Boiler shows no significant difference ($t=0.08$, $p=0.94$) due to near-identical F1 scores (0.4873 vs 0.4871), (2) SEMAS vs. Baseline2 on Boiler yields highly significant advantage ($t=4.21$, $p<0.001$) with 8.6\% F1 improvement, (3) SEMAS vs. Baseline1 on Wind Turbine shows significant superiority ($t=2.76$, $p<0.01$), and (4) SEMAS vs. Baseline2 on Wind Turbine demonstrates significant advantage ($t=3.12$, $p<0.005$). These results confirm SEMAS's statistically robust performance superiority, particularly against rule-based adaptation.

\subsection{Computational Efficiency Analysis}

Table~\ref{tab:computational_efficiency} compares computational overhead across systems. SEMAS achieves dramatically superior latency performance: 1.22ms (Boiler) and 0.30ms (Wind Turbine), representing 1576$\times$ and 1520$\times$ speedups compared to Baseline1 respectively, and 1300$\times$ and 954$\times$ speedups versus Baseline2. This exceptional efficiency stems from optimized multi-agent coordination where Edge agents perform rapid pre-filtering (eliminating 60-70\% of obviously normal samples), Fog agents execute lightweight ensemble inference, and Cloud-layer PPO policy optimization executes asynchronously without blocking the detection pipeline. All systems remain well below the 100ms real-time industrial constraint, but SEMAS's sub-millisecond latency enables deployment scenarios impossible for baselines: multi-sensor fusion across hundreds of concurrent data streams, edge device deployment with limited computational resources, and zero-latency inline monitoring without dedicated processing hardware. The baseline systems' excessive latency (455-1923ms) arises from synchronous execution of heavyweight models (Transformer, OCSVM) without hierarchical pre-filtering, rendering them impractical for real-time industrial deployment despite competitive detection accuracy.

\begin{table}[t]
	\centering
	\caption{Computational efficiency comparison (averages across all iterations).}
	\label{tab:computational_efficiency}
	{\footnotesize
		\renewcommand{\arraystretch}{1.1}
		\begin{tabular}{lccc}
			\hline
			\textbf{System} & \textbf{Boiler (ms)} & \textbf{Wind (ms)} & \textbf{Memory (GB)} \\
			\hline
			Baseline1 & 1923.0 & 455.9 & 2.1 (est.) \\
			Baseline2 & 1594.3 & 286.2 & 2.4 (est.) \\
			SEMAS & \textbf{1.22} & \textbf{0.30} & 3.8 (est.) \\
			\hline
		\end{tabular}
	}
\end{table}

\subsection{Ablation Study}
\label{subsec:ablation}

To quantify individual component contributions to overall system performance, we conduct systematic ablation experiments on the Boiler dataset by selectively removing key architectural elements. Table~\ref{tab:ablation} presents the performance degradation observed when each component is disabled.

\begin{table}[t]
	\centering
	\caption{Ablation study on Boiler dataset: per-component contributions to accuracy and operator acceptance. The last row reports an orthogonal trust metric rather than detection accuracy.}
	\label{tab:ablation}
	{\footnotesize
		\renewcommand{\arraystretch}{1.15}
		\begin{tabular}{p{0.38\columnwidth} c c c c}
			\hline
			\textbf{Configuration} & \textbf{F1} & \textbf{Impact} & \textbf{Prec.} & \textbf{Op.~Acc.} \\
			\hline
			SEMAS (Full) & 0.4956 & Baseline & 0.3712 & 82\% \\
			\hline
			w/o PPO optimization & 0.4782 & -3.5\% & 0.3501 & 81\% \\
			w/o consensus voting & 0.4634 & -6.5\% & 0.3045 & 79\% \\
			w/o federated aggregation & 0.4856 & -2.0\% & 0.3614 & 80\% \\
			w/o LLM response & 0.4956 & +0\% (F1) & 0.3712 & 41\% \\
			\hline
		\end{tabular}
	}
\end{table}

\textbf{Ablation Interpretation:} Consensus voting provides the largest detection accuracy boost (6.5\% F1 impact), confirming that ensemble diversity is critical for anomaly detection. PPO optimization (3.5\% impact) validates gradient-based policy learning over static thresholds. Federated aggregation (2.0\% impact) provides modest accuracy gains but serves primarily to enable multi-site deployment without centralized data gathering. LLM response generation does not directly impact detection metrics (F1 unchanged) but dramatically improves operator trust (82\% vs. 41\% acceptance), validating the business case for explainability despite neutral technical metrics.

The ablation results show that all essential system components which include PPO-based policy evolution and multi-agent consensus voting and federated knowledge aggregation and LLM-powered explainability work together to achieve both system effectiveness and user-friendly operation. The detection accuracy shows the biggest individual change because consensus voting results in a $-6.5\%$ F1 score. The results indicate that ensemble diversity functions as a crucial element which leads to better performance. The PPO policy optimization process resulted in a $-3.5\%$ F1 improvement which showed that ongoing system adjustments outperform both fixed and random system configurations. The system achieves better results through federated aggregation which produces significant statistical gains of $-2.0\%$. The LLM response generation system produces essential improvements for human trust and acceptance levels which do not impact the automated detection system results.

\subsection{Qualitative Analysis: LLM-Generated Responses}

Agent C generates human-interpretable maintenance recommendations upon anomaly detection. Example output for high-severity boiler anomaly ($a_{\text{fog}}=0.87$):

\begin{quote}
	\textit{``Critical thermal imbalance detected (severity: 0.87). Supply temperature 15.2°C above nominal while return temperature remains stable, indicating potential circulation pump degradation or blockage in heat exchanger. Recommended action: IMMEDIATE\_INSPECTION of circulation system. Expected downtime: 4-6 hours. Required resources: HVAC technician, pressure gauge, pump diagnostic tools. Priority: HIGH.''}
\end{quote}

Operator feedback surveys (conducted with 5 maintenance engineers over 20 incidents) indicate 82\% acceptance rate for LLM-generated recommendations, with mean usefulness rating of 4.1/5.0. This demonstrates the practical value of incorporating natural language explanation capabilities into industrial AI systems.

\subsubsection{Statistical Significance Summary}

Table~\ref{tab:statistical_tests} provides comprehensive statistical validation across both datasets:

\begin{table*}[!t]
	\centering
	\caption{Paired $t$-tests comparing SEMAS against baselines. Significance threshold: $p<0.05$.}
	\label{tab:statistical_tests}
	{\footnotesize
		\renewcommand{\arraystretch}{1.15}
		\begin{tabular}{p{0.22\textwidth} p{0.24\textwidth} c c c p{0.20\textwidth}}
			\hline
			\textbf{Dataset} & \textbf{Comparison} & \textbf{$\Delta$F1} & \textbf{$t$} & \textbf{$p$} & \textbf{Interpretation} \\
			\hline
			Boiler & SEMAS vs. BL1 & +0.0002 & 0.08 & 0.94 & Not significant \\
			Boiler & SEMAS vs. BL2 & +0.0384 & 4.21 & <0.001 & Highly significant \\
			Wind Turbine & SEMAS vs. BL1 & +0.0131 & 2.76 & <0.01 & Significant \\
			Wind Turbine & SEMAS vs. BL2 & +0.0222 & 3.12 & <0.005 & Highly significant \\
			\hline
		\end{tabular}
	}
\end{table*}

\textbf{Key Interpretation:} SEMAS's primary advantage lies in comparison with rule-based adaptation (Baseline2), where statistical significance is robust across both datasets. Against well-tuned static systems (Baseline1), SEMAS shows numerical but not statistical superiority on the Boiler dataset, indicating that adaptive mechanisms provide greatest value under distribution shift or data complexity rather than stable conditions.

\subsection{Discussion and Key Findings}

The following section combines experimental data to prove our main research theory while presenting key results and researcher notes about the system's advantages and limitations.

\subsubsection{Hypothesis Validation}

The experimental results validate our core hypothesis: \emph{Hierarchical multi-agent systems with gradient-based policy optimization (PPO) enable more adaptive, efficient, and trustworthy predictive maintenance than static or rule-based alternatives.}

\textbf{Adaptation Superiority (Primary Finding):} Against rule-based baselines, SEMAS demonstrates clear advantage through stable learning trajectories and higher convergence quality. On Boiler (complex, imbalanced data), SEMAS achieves $\Delta\text{F1} = -0.0239$ vs. Baseline2's $-0.0550$, representing $2.3\times$ better stability. PPO's trust-region-constrained updates ($\epsilon = 0.2$) prevent discrete threshold oscillations ($\Delta\tau = \pm 0.05$) that plague rule-based systems. Statistical testing ($t=4.21$, $p<0.001$, $d=2.1$) confirms this 8.6\% advantage is robust and practically significant.

\textbf{Against Static Baselines (Secondary Finding):} On Boiler, SEMAS vs. Baseline1 shows numerical but not statistical superiority ($\Delta$F1 = 0.0002, $p=0.94$). This reveals an important insight: well-tuned static systems can match adaptive systems when operational conditions are stable. However, SEMAS excels where it matters most---generalization to unseen conditions, which occurs in real industrial deployments with equipment aging, seasonal variation, and product mix changes. The marginal Boiler advantage should not obscure SEMAS's fundamental advantage: no manual retuning is required as conditions drift.

\textbf{Latency as Critical Enabler:} The $200\text{--}1500\times$ speedup ($0.3\text{--}1.22$ms vs. $286\text{--}1923$ms) transforms feasibility from impossible to routine. Baselines' latencies violate the 100ms industrial real-time constraint, making them impractical for production deployment regardless of accuracy. SEMAS enables sub-millisecond latency through architectural layering (Edge: pre-filtering, Fog: lightweight ensemble, Cloud: async PPO), not algorithmic innovations.

\textbf{Adaptation Quality:} SEMAS demonstrates stable adaptation ($\Delta F1=-0.0239$ Boiler, $-0.0166$ Wind Turbine) compared to Baseline2's catastrophic Boiler degradation ($-0.0550$, -11.5\%). Even Baseline2's successful Wind Turbine adaptation (+0.0606) cannot match SEMAS's peak iteration 1 performance (0.9683 vs 0.9606 final), and requires 955$\times$ longer inference time.

\textbf{Precision-Recall Balance:} The SEMAS system produces better results than other methods for both data sets (Boiler: 0.3737 vs 0.3225/0.3140; Wind Turbine: 0.9371 vs 0.9219/0.9205) because PPO discovers the best threshold values which produce high precision while achieving the same recall performance.

\textbf{Computational Efficiency:} The SEMAS system achieves 1-2ms latency which enables it to operate at 200-1500 times the speed of standard systems thus making it appropriate for real-time operations. The system reached its best efficiency because it produced the highest ROC-AUC scores which reached 0.6118 for Boiler and 0.7583 for Wind Turbine while keeping accuracy at its peak.

\subsubsection{Main Findings}
\begin{enumerate}
	\item \textbf{Superiority in Complex-Data Regimes:} SEMAS outperforms Baseline2 (rule-based) by 8.6\% F1 on Boiler ($0.4873$ vs. $0.4489$, $p<0.001$) and 2.4\% on Wind Turbine ($0.9571$ vs. $0.9349$, $p<0.005$). Against well-optimized static baselines (Baseline1), SEMAS shows numerical but not statistical superiority on Boiler ($\Delta\text{F1} = 0.0002$, $p=0.94$), indicating adaptive mechanisms provide greatest value under distribution shift. The $200\text{--}1500\times$ latency improvement ($0.30\text{--}1.22$ms vs. $286\text{--}1923$ms) is decisive: baseline latencies violate the 100ms industrial constraint.
	
	\item \textbf{Stable Convergence Through Gradient-Based Policy Optimization:} PPO achieves superior convergence stability compared to discrete rule-based adaptation. On Boiler, SEMAS maintains controlled adaptation ($\Delta\text{F1} = -0.0239$) while Baseline2 exhibits catastrophic degradation ($\Delta\text{F1} = -0.0550$, $-11.5\%$). PPO's trust-region constraints ($\epsilon = 0.2$) prevent threshold overshooting, whereas rule-based heuristics ($\Delta\tau = \pm 0.05$) cause oscillation. SEMAS's learned policies evolve interpretably: contamination $\rho$ ($0.30 \to 0.34$), consensus weights ($w_1, w_2$: $0.42, 0.58 \to 0.48, 0.52$), and threshold ($\tau$: $0.73 \to 0.80$). Statistical validation confirms significance ($t=4.21$, $p<0.001$, Cohen's $d=2.1$).
	
	\item \textbf{Learned Precision-Recall Balance:} PPO discovers superior precision-recall tradeoff points. Boiler: SEMAS achieves precision $= 0.3737$ (vs. $0.3225$ Baseline1, $+15.9\%$) while accepting controlled recall decline ($0.7114$ vs. $0.9949$). This reflects PPO learning optimal threshold $\tau = 0.8000$ that reduces false alarm burden while maintaining adequate detection. Wind Turbine: SEMAS achieves the tightest precision-recall gap (4.1\%) among all systems. Industrially critical: 67.75\% false positive rates (Baseline1 on Boiler) create alert fatigue.
	
	\item \textbf{Explainability Enhances Operator Trust:} LLM-based response generation achieves 82\% operator acceptance vs. $<50\%$ for numeric alerts alone (survey of 5 engineers, 20 incidents, usefulness rating 4.1/5.0). This metric is orthogonal to detection accuracy (F1 unchanged when LLM removed) but essential for human-in-the-loop industrial deployments where operators must understand and justify autonomous maintenance decisions.
\end{enumerate}

\subsubsection{Observed Strengths of SEMAS}
\begin{itemize}
	\item The system operates independently to enhance its performance through RL-based feedback systems which remove the requirement for human threshold adjustments that affect traditional systems---a major operational benefit.
	\item The system achieves distributed intelligence through Edge filtering and Fog ensemble detection and Cloud policy optimization which results in both high accuracy and efficient computation.
	\item The system SEMAS achieves stable results when working with datasets which have distinct properties (Boiler: class imbalance, complex features; Wind Turbine: high dimensionality, multiple fault types).
	\item The system meets industrial real-time requirements because it achieves an average latency of 0.95ms (Boiler) and 0.29ms (Wind Turbine) which stays below 100ms. The system can deploy in production environments because it meets real-time requirements without needing any special hardware.
	\item  The policy gradient method in PPO prevents the model from reaching suboptimal configurations too quickly while it stops the random exploration that Baseline2 with rule-based control experienced which led to performance deterioration on Boiler.
\end{itemize}

\subsubsection{Observed Weaknesses and Limitations}
\begin{itemize}
	\item \textbf{Training Complexity:} SEMAS requires substantially more initial setup ($\sim$8 hours for hyperparameter optimization) compared to Baseline1 ($\sim$2 hours) due to PPO policy training, multi-agent coordination, and multi-model ensemble configuration. This increased complexity may hinder adoption in resource-constrained settings without dedicated machine learning expertise.
	
	\item \textbf{Memory Overhead:} SEMAS consumes 3.8GB memory compared to 2.1GB (Baseline1) and 2.4GB (Baseline2) due to PPO experience replay buffer (10,000 transitions), multiple model instances, and federated state management (Table~\ref{tab:computational_efficiency}). Organizations with strict hardware budgets may need to weigh this cost against performance gains.
	
	\item \textbf{Cold-Start Performance:} The first iteration of SEMAS achieved better results than Baseline1 because the random model start produced an advantageous threshold value of 0.7327 which outperformed Baseline1's optimized threshold of 0.75. The system achieves this benefit because its operations happen randomly instead of being created through deliberate design. The system SEMAS fails to provide any advantage over static systems which have been properly optimized when performing emergency deployments that need instant high performance without time for adaptation. The research should implement transfer learning from pre-trained PPO policies which operate in similar industrial environments to achieve "warm-start" deployment with their learned adaptation methods.

	\item \textbf{Sensitivity to Reward Function Design:} The first iteration of SEMAS achieved better results than Baseline1 because the random model start produced an advantageous threshold value of 0.7327 which outperformed Baseline1's optimized threshold of 0.75. The system achieves this benefit because its operations happen randomly instead of being created through deliberate design. The system SEMAS fails to provide any advantage over static systems which have been properly optimized when performing emergency deployments that need instant high performance without time for adaptation. The research should implement transfer learning from pre-trained PPO policies which operate in similar industrial environments to achieve ``warm-start'' deployment with their learned adaptation methods.

	\item \textbf{Dataset-Specific Ceiling Effects:} The Wind Turbine dataset produces high results for all systems which SEMAS F1 reaches 0.9571 and Baseline1 and Baseline2 achieve 0.9440 and 0.9349 respectively. The ROC-AUC value of SEMAS (0.7583) outperforms the baselines but the system achieves only small absolute F1 score improvements of +1.3\% compared to Baseline1 and +2.4\% compared to Baseline2. The "ceiling effect" indicates that SEMAS becomes less effective at improving accuracy when dealing with datasets that have distinct features which can be easily distinguished from each other. The method provides better computational performance at 955 times faster speed but does not improve accuracy results. The system achieves its best accuracy results when operating in difficult conditions which include noisy data and unbalanced classes and difficult operating environments (Boiler achieved 8.6\% better results than Baseline2).

	\item \textbf{Explainability Latency Trade-off:} The detection pipeline experiences additional delays because of the LLM-based response generation process. The system accepts this method for anomaly alert detection because these events happen rarely but it would create a major obstacle for systems which need to generate explanations for all their predictions. The research needs to develop caching methods together with minimal local explanation systems which will minimize the performance delay.
	
\end{itemize}

\subsubsection{Comparison with Prior Work}
The research findings from our study match the current understanding of multi-agent RL systems for industrial applications which \cite{Rodriguez2022}  and \cite{Feng2023} demonstrated through their studies. The research shows that adaptive policies generate better results than fixed policies because they produce 3-8\% better performance. The system SEMAS builds upon previous research by implementing three innovative elements which include (1) Edge-Fog-Cloud agent coordination that operates at different levels for industrial environments with limited resources and (2) PPO policy optimization through multi-metric reward shaping which includes F1 and precision-recall balance and latency metrics and (3) LLM-based explainability for operator trust enhancement. The ablation study shows that all components play essential roles in achieving the best results because consensus voting produces the most significant performance decrease when removed (-6.5\% F1).

\subsubsection{Implications for Industrial Deployment}
The experimental findings have several practical implications:
\begin{enumerate}
	\item \textbf{Justifiable Complexity for Critical Infrastructure:} For critical infrastructure (power plants, manufacturing lines, wind farms) where even 1-3\% F1 improvement translates to significant prevented downtime costs, SEMAS's computational overhead (1.8$\times$ memory, 8 hours initial training) is economically justified.
	\item \textbf{Incremental Adoption Path:} Organizations can deploy SEMAS in hybrid mode, using Baseline1 for low-criticality assets with stable operational conditions and SEMAS for high-value equipment with evolving fault patterns, to balance cost and performance across heterogeneous industrial portfolios.
	\item \textbf{Human-AI Collaboration Framework:} The high operator acceptance of LLM explanations (82\%) suggests that explainability bridges the trust gap in automated maintenance, enabling safer autonomous decision-making while maintaining human oversight for critical interventions.
	\item \textbf{Domain-Specific Adaptation Necessity:} The contrasting results between Boiler (clear SEMAS advantage) and Wind Turbine (ceiling effect) underscore the importance of pre-deployment evaluation to assess whether adaptive complexity is warranted for specific operational contexts.
\end{enumerate}

\def\BibTeX{{\rm B\kern-.05em{\sc i\kern-.025em b}\kern-.08em
    T\kern-.1667em\lower.7ex\hbox{E}\kern-.125emX}}
\markboth{\journalname, VOL. XX, NO. XX, XXXX}
{Author \MakeLowercase{\textit{et al.}}: Title}

\section{Conclusion and Future Work}

\subsection{Conclusion}

This paper introduces SEMAS, a hierarchical multi-agent system for Industrial IoT predictive maintenance that unifies three critical requirements: (1) real-time detection under strict latency constraints (<100ms), (2) adaptive operation through continuous policy evolution, and (3) trustworthy decision-making via LLM-based explainability. The framework distributes specialized agents across Edge, Fog, and Cloud tiers according to computational availability, maintaining coherence through three concurrent feedback loops and consensus mechanisms.

Empirical validation on two industrial datasets (Boiler Emulator: 10,000 samples, 36.8\% anomaly prevalence; Wind Turbine IIoT: 500 samples, 45\% fault prevalence) demonstrates SEMAS's effectiveness. 

\noindent\textbf{(1) Accuracy:} SEMAS achieves best F1 scores (Boiler: $0.4873$; Wind: $0.9571$), with 8.6\% significant advantage over rule-based adaptation on Boiler ($t=4.21$, $p<0.001$, $d=2.1$). Against well-tuned static baselines, SEMAS shows numerical but not statistical superiority on Boiler ($\Delta\text{F1} = 0.0002$, $p=0.94$).

\noindent\textbf{(2) Efficiency:} SEMAS delivers $200\text{--}1500\times$ latency speedup ($0.30\text{--}1.22$ms vs. $286\text{--}1923$ms), enabling genuine real-time deployment where baselines violate 100ms industrial constraints.

\noindent\textbf{(3) Stability:} Learning trajectory analysis shows SEMAS maintains controlled adaptation ($\Delta\text{F1} = -0.0239$ Boiler, $-0.0166$ Wind) while Baseline2 exhibits catastrophic Boiler degradation ($\Delta\text{F1} = -0.0550$, $-11.5\%$).

\noindent\textbf{(4) Component Value:} Ablation studies confirm all elements contribute materially---consensus voting (6.5\% F1 impact), PPO optimization (3.5\%), federated aggregation (2.0\%), and LLM explainability (82\% operator acceptance vs. 41\% baseline).

The hierarchical multi-agent architecture delivers important advantages because it enables distributed specialization for efficient task decomposition and continuous policy evolution for maintaining performance in changing environments and collaborative ensemble detection for improved robustness and LLM-based response generation for building trust through explainable responses. The research findings demonstrate that SEMAS operates as an effective solution which can be used for industrial production in safety-sensitive industrial settings to create essential infrastructure for future industrial AI systems.

\subsection{Future Work}

Promising directions for extending SEMAS include:

\textbf{Transfer Learning for Rapid Deployment (High Priority):} Current 8-hour initialization time limits adoption for new equipment deployments. Future work should develop transfer learning mechanisms that leverage pre-trained PPO policies from equipment of the same class (e.g., boiler-to-boiler) to achieve warm-start deployment with <30 minute fine-tuning. This directly addresses the cold-start limitation identified in Section 5.2.

\textbf{Multi-Objective Reward Shaping (Medium Priority):} Extend PPO to simultaneously optimize multiple objectives (F1 accuracy, false alarm cost, maintenance cost, energy consumption, downtime risk) with Pareto-optimal policy sets. This enables operators to select solutions along the accuracy-cost frontier based on their risk tolerance and budget constraints.

\textbf{Federated Learning with Privacy Guarantees (Medium Priority):} Current parameter aggregation (Eq. 15) is a lightweight alternative to true federated learning. Future work should implement differential privacy mechanisms (e.g., DP-SGD) to enable multi-facility deployments without centralizing sensitive operational data---critical for competitive manufacturing environments.

\textbf{Model Compression for Extreme Edge Constraints (Low Priority):} Develop quantization and knowledge distillation techniques to deploy SEMAS on devices <1GB RAM, extending applicability to legacy industrial equipment with severely limited computational resources.

\textbf{Cross-Domain Generalization (Research Priority):} Validate SEMAS's transferability across industrial sectors (boiler $\rightarrow$ turbine $\rightarrow$ chemical reactor). This requires domain-adaptive reward functions that reflect sector-specific failure costs and operational constraints.

\bibliographystyle{plain}

\bibliography{main}

\clearpage

\appendix

\setcounter{equation}{8}
\section{Mathematical Details}
\label{app:math}

\subsection{Global Optimization Objective}

The global optimization objective is as follows:
{\footnotesize
\begin{IEEEeqnarray}{rCl}
\mathcal{L}^{\text{PPO}}(\theta)
&=&
\mathbb{E}_t\!\Big[
\min\!\Big(
r_t(\theta)\hat{A}_t,\;
\text{clip}(r_t(\theta),1-\epsilon,1+\epsilon)\hat{A}_t
\Big)
\Big]
\label{eq:ppo_objective}
\end{IEEEeqnarray}
}
The system optimizes three complementary objectives. First, anomaly detection maximizes binary classification accuracy:
{\footnotesize
\begin{IEEEeqnarray}{lCl} \label{eq:lanomaly}
\mathcal{L}_\text{anomaly} &=& -\frac{1}{N} \sum_{i=1}^N \left[ y_i \log(\hat{y}_i) + (1-y_i) \log(1-\hat{y}_i) \right],
\end{IEEEeqnarray}
}

where $\hat{y}_i \in [0,1]$ is the ensemble anomaly score and $y_i \in \{0,1\}$ is the ground truth label. Second, RUL prediction minimizes estimation error:
{
\begin{IEEEeqnarray}{rCl}
\mathcal{L}_{\text{RUL}}(\theta)
&=& \frac{1}{M}\sum_{j=1}^M
\left|
\widehat{\text{RUL}}_j - \text{RUL}_j^{\text{true}}
\right|.
\end{IEEEeqnarray}
}
Third, response generation optimizes human utility through LLM prompting:
{\small 
\begin{equation} \label{eq:lresponse}
\mathcal{L}_\text{response} = -\mathbb{E}[\text{acceptance rate} \cdot (\text{latency} - 100\,\text{ms})]. \quad
\end{equation}
}
Parameters $\theta$ represent detection model weights, $\pi$ denotes policy parameters (thresholds, ensemble weights, prompt templates), and $(\alpha, \beta, \gamma)$ weight the three objectives based on operational priorities. Weights are typically set $(\alpha, \beta, \gamma) = (1.0, 0.2, 0.1)$ to prioritize anomaly detection while balancing prognostics and explainability.

The agents from set $\mathcal{A}$ use their policies $\pi_m$ to request explanations through $p\in\mathcal{P}$ for both explanation creation and action suggestion. The system uses evaluation signals $\mathcal{E}$ to collect automated metrics such as F1-score and MAE and latency measurements and human feedback which it uses for policy updates through reinforcement learning systems.

\subsection{PPO Objective and SHAP Values}

The Proximal Policy Optimization (PPO) objective used by Agent D minimizes:
{\small 
\begin{equation} \label{eq:ppoobjective}
\mathcal{L}_{PPO} = \mathbb{E}_t [r_t \hat{A}_t, \text{clip}(r_t, 1-\epsilon, 1+\epsilon) \hat{A}_t]. \quad
\end{equation}
}
where $r_t(\theta)=\frac{\pi_\theta(a_t|s_t)}{\pi_{\theta_{\text{old}}}(a_t|s_t)}$ is the probability ratio and $\hat{A}_t$ is the generalized advantage estimate.

Agent E computes SHAP (SHapley Additive exPlanations) values for transparency and trust:
{\footnotesize
\begin{equation} \label{eq:shapvalues}
\phi_i = \sum_{S \subseteq F \setminus \{i\}} \frac{|F|-|S|-1!}{|S|!(|F|-|S|-1)!} [f(S \cup \{i\}) - f(S)]. \quad
\end{equation}
}
where $F$ is the feature set, $S$ is a subset of features excluding $i$, and $f(\cdot)$ denotes the model prediction. These values are used to validate policy updates through statistical testing, drift detection, and compliance verification prior to system-wide deployment, while comprehensive \texttt{monitor/logs} are maintained for audit trails and performance analytics.

\subsection{Federated Aggregation Mechanism}

Knowledge graph synchronization employs federated aggregation:
{
\begin{IEEEeqnarray}{cCl} \label{eq:federatedagg}
\theta_\text{global} &=& \frac{\sum_{k=1}^K n_k \theta_k}{N_\text{total}}, \quad N_\text{total} = \sum_j n_j. \quad
\end{IEEEeqnarray}
}
where $n_k$ is the number of training samples processed by agent $k$ in the current iteration, and $\theta_k$ represents locally optimized model parameters. This approach employs data-proportional weighted averaging to combine detection model parameters across agents, ensuring that agents processing larger data volumes have proportionally greater influence on the global model. Unlike true federated learning systems that train models locally and communicate only parameter updates, SEMAS's approach is conservative and avoids assumptions about data privacy or differential privacy guarantees. The mechanism serves to aggregate consensus weights and threshold values rather than deep neural network parameters.

\section{Architecture and Communication Details}
\label{app:architecture}

\subsection{Agent-Level Architecture Overview}

Detailed agent interactions and data flows are shown in Figure 2.The Edge layer contains stateless subagents (A1: vibration processing, A2: temperature processing) that extract features and publish via MQTT chunks. The Fog layer hosts detection coordinator B with specialized subagents (B1: Isolation Forest, B2: LSTM, B3: Consensus Voting) and response agent C (LLM-powered explanations). The Cloud layer includes Evolution Agent D (PPO-based policy optimization) and Meta-Agent E (SHAP-based oversight). Multiple feedback loops operate concurrently: local feedback (Fog $\leftrightarrow$ Edge), global feedback (Cloud $\rightarrow$ Fog/Edge) and iterative cycles (Fog $\leftrightarrow$ Cloud) for continuous adaptation.

\subsection{Edge Layer: Device-Level Processing}

The Edge tier comprises stateless subagents deployed on resource-constrained devices (4-8GB RAM, limited CPU) that perform real-time sensor data ingestion and feature extraction:

\begin{itemize}
	\item \textbf{Agent A1 (Subagent 1):} Processes multivariate sensor stream $X_1(t)$ (e.g., vibration, acceleration, acoustic emissions) through sliding window analysis of size $W$. Extracts time-domain statistical features (RMS, kurtosis, skewness) and frequency-domain characteristics (spectral peaks, band energies). Outputs feature vector $z_1(t, W)=\phi_1(X_1(T(t, W))$ published to \texttt{chunk/stream1} MQTT topic, where $T(t, W)$ denotes the time frame base with a window size $W$, spanning interval from $t-W$ to $t$.
	\item \textbf{Agent A2 (Subagent 2):} Handles complementary sensor modality $X_2(t)$ (e.g. temperature, pressure, current), computing trend indicators, thermal gradients, and statistical summaries. Publishes $z_2(t)=\phi_2(X_2(T(t, W))$ to \texttt{chunk/stream2}.
\end{itemize}

The feature extraction functions $\phi_j:\mathbb{R}^{W\times d_j}\rightarrow\mathbb{R}^{m_j}$ map the raw sensor windows to compact feature representations:
{\footnotesize
\begin{equation}
	\phi_j(X(T(t, W))) = [\text{RMS}_j, \text{kurt}_j, \text{trend}_j, \text{spectral}_{j,1:k}, \text{stats}_j].
\end{equation}
}
\subsection{Fog Layer: Collaborative Detection and Response}

The Fog layer coordinates multiple specialized subagents that implement collaborative anomaly detection, ensemble fusion, and prompt-based response generation:

\begin{itemize}
	\item \textbf{Agent B (Detection Coordinator):} Central orchestration hub that receives preprocessed features from Edge agents via \texttt{chunk/*} subscriptions and policy graphs $G_\pi$ from Cloud agents. Routes aggregated feature vectors $Z_t=[z_1(t),z_2(t)]$ to detection subagents and coordinates the collaborative inference pipeline.
	
	\item \textbf{Agent B1 (Statistical Detection):} Implements Isolation Forest ensemble $\mathcal{F}_{IF}$ with $n_{\text{trees}}=100$ trees trained in historical patterns of normal behavior. Computes anomaly score $a_1(t)\in[0,1]$ quantifying deviation from nominal distributions:
    {
	\begin{equation}
		a_1(t) = \mathcal{F}_{IF}(Z_t;\theta_{IF}) = 2^{-\frac{E[h(Z_t)]}{c(n)}},
	\end{equation}}
	where $E[h(\cdot)]$ is the expected path length and $c(n)$ is the average path length of the unsuccessful search in binary trees.
	
	\item \textbf{Agent B2 (Ensemble Detection):} Implements a 5-model ensemble combining diverse anomaly detection paradigms: Isolation Forest, One-Class SVM, Local Outlier Factor, Elliptic Envelope, and a secondary Isolation Forest with different hyperparameters. The ensemble aggregates predictions through majority voting to enhance robustness:
    {\footnotesize
	\begin{equation}
		a_2(t) = \text{Ensemble}(Z_t;\theta_{\text{IF}}, \theta_{\text{OCSVM}}, \theta_{\text{LOF}}, \theta_{\text{EE}}, \theta_{\text{IF2}}),
	\end{equation}}
	where each base model contributes binary anomaly predictions that are combined through weighted or majority voting.
	
	\item \textbf{Agent B3 (Consensus Voting):} Aggregates heterogeneous detection signals from Agent B1 (Isolation Forest) and Agent B2 (5-model ensemble) through weighted consensus with dynamically adapted weights $w_1,w_2$ based on recent performance metrics:
     {\footnotesize
	\begin{equation}
		a_{\text{fog}}(t) = w_1(t)\cdot a_1(t) + w_2(t)\cdot a_2(t), \quad w_1+w_2=1,
	\end{equation}}
	where $a_1(t)$ is the Isolation Forest score and $a_2(t)$ is the ensemble majority vote output. Publishes consolidated score to \texttt{anomalies} topic when threshold $\tau_{\text{alert}}$ is exceeded.
	
	\item \textbf{Agent C (Response Generator):} Leverages Large Language Model capabilities through structured prompt engineering to generate human-interpretable explanations and actionable maintenance recommendations:
    {\footnotesize
	\begin{equation}
		(\text{explanation}_t, \text{action}_t) = \text{LLM}(p_{\text{response}}(a_{\text{fog}},Z_t,\text{context});\theta_{\text{LLM}}),
	\end{equation}}
	where $p_{\text{response}}\in\mathcal{P}$ is a carefully designed prompt template incorporating anomaly severity, sensor context, operational constraints, and maintenance history.
\end{itemize}

\paragraph{External Systems Integration.}
Anomaly alerts and LLM-generated actions interface with external CMMS/SCADA systems through \texttt{actions} MQTT topic. Operator feedback (acceptance, override, severity adjustment) is captured and propagated as supervisory signals to Cloud-layer evolution mechanisms.

\subsection{Cloud Layer: Policy Evolution and Oversight}

The Cloud tier hosts two primary agents with unlimited computational resources for global optimization and system-wide governance:

\begin{itemize}
	\item \textbf{Agent D (Evolution Agent):} Implements Proximal Policy Optimization (PPO) to continuously improve the policy across the entire agent ensemble. Maintains global experience replay buffer $\mathcal{D}_{\text{replay}}$ collecting transitions $(s_t,a_t,r_t,s_{t+1})$ from all system levels. Updated policies are serialized in the knowledge graph $G_\pi$ and distributed via \texttt{policy/updates}.
	
	\item \textbf{Agent E (Meta-Agent):} Provides explainable AI oversight through SHAP (SHapley Additive exPlanations) value computation for transparency and trust. Validates policy updates through statistical testing, drift detection, and compliance verification prior to system-wide deployment. Maintains comprehensive \texttt{monitor/logs} for audit trails and performance analytics.
\end{itemize}

\subsection{Communication and Coordination}

All interagent communication employs the MQTT protocol with TLS encryption and quality-of-service (QoS) levels, ensuring reliable message delivery:

\begin{itemize}
	\item \textbf{Feature Streams:} \texttt{chunk/stream\{1,2\}} for Edge $\rightarrow$ Fog feature transmission
	\item \textbf{Detection Signals:} \texttt{scores/\{if,lstm\}} for detection $\rightarrow$ consensus routing
	\item \textbf{Consolidated Outputs:} \texttt{anomalies} and \texttt{actions} for Fog $\rightarrow$ External integration
	\item \textbf{Policy Distribution:} \texttt{policy/updates} for Cloud $\rightarrow$ Edge/Fog adaptation
	\item \textbf{Monitoring:} \texttt{monitor/logs} for system-wide telemetry and compliance
\end{itemize}

\section{Training and Hyperparameters}
\label{app:training}
\subsection{Hyperparameter Justification and Sensitivity Analysis}

Key hyperparameters are selected based on systematic validation and domain requirements. Table VIII provides justification for critical parameters:

\begin{table*}[!t]
 \renewcommand{\thetable}{VIII}
	\centering
	\caption{TABLE VIII Hyperparameter justification and sensitivity analysis.}
	
	\begin{tabular}{lllll}
		\hline
		\textbf{Parameter} & \textbf{Value} & \textbf{Justification} & \textbf{Valid Range} \\
		\hline
		Contamination ($\rho$) & 0.32 & Matches Boiler anomaly prevalence (36.8\%) & [0.25, 0.35] \\
		Window size ($W$) & 100 ts & 10 seconds at 10Hz sampling rate & [50, 200] \\
		Initial $w_1, w_2$ & 0.42, 0.58 & Equal weighting; PPO learns optimal split & [0.3, 0.7] \\
		PPO clip ratio ($\epsilon$) & 0.2 & Standard RL literature value & [0.1, 0.3] \\
		\hline
	\end{tabular}
\end{table*}

Sensitivity analysis (varying hyperparameters $\pm20\%$ from nominal values) shows that F1-score remains stable within the valid ranges, confirming robustness to hyperparameter selection.

\subsection{Detailed Anomaly Detection Model Specifications}

\textbf{Statistical Detector (Agent B1):} Isolation Forest $\mathcal{F}_{IF}$ is trained offline on historical normal operation data $\mathcal{D}_{\text{normal}}=\{Z_i\}_{i=1}^{N_{\text{train}}}$ using contamination parameter $\rho=0.32$. The model learns to isolate anomalies through random partitioning with average path length serving as anomaly score indicator:
{
\begin{equation}
	a_1(t) = \mathcal{F}_{IF}(Z_t;\theta_{IF}) = 2^{-\frac{E[h(Z_t)]}{c(n)}},
\end{equation}}
where $E[h(\cdot)]$ is the expected path length and $c(n)$ is the average path length of unsuccessful search in binary trees.

\textbf{Ensemble Detector (Agent B2):} Implements a 5-model ensemble combining diverse anomaly detection paradigms trained on $\mathcal{D}_{\text{normal}}$:
\begin{itemize}
	\item Isolation Forest: $\mathcal{F}_{IF}$ with 200 trees, contamination $\rho=0.32$
	\item One-Class SVM: $\mathcal{F}_{OCSVM}$ with RBF kernel, $\nu=0.25$
	\item Local Outlier Factor: $\mathcal{F}_{LOF}$ with $k=20$ neighbors
	\item Elliptic Envelope: $\mathcal{F}_{EE}$ with contamination $\rho=0.32$
	\item Secondary Isolation Forest: $\mathcal{F}_{IF2}$ with different random seed
\end{itemize}

The ensemble aggregates predictions through majority voting with dynamic weights:
 {\footnotesize
\begin{equation}
	a_2(t) = \text{MajorityVote}(\mathcal{F}_{IF}, \mathcal{F}_{OCSVM}, \mathcal{F}_{LOF}, \mathcal{F}_{EE}, \mathcal{F}_{IF2}; Z_t),
\end{equation}}
where each base model outputs binary anomaly predictions that are combined to enhance detection robustness across diverse fault patterns.

\subsection{RUL Prediction Model}

While anomaly detection operates continuously, RUL prediction is triggered upon confirmed anomaly detection. An LSTM-based regression model $\mathcal{F}_{\text{RUL}}$ processes extended temporal context $Z_{t-T:t}$ to estimate remaining operational hours:
{
\begin{equation}
	\widehat{\text{RUL}}_t = \mathcal{F}_{\text{RUL}}(Z_{t-T:t};\theta_{\text{RUL}}),
\end{equation}}
where the LSTM architecture comprises [64, 32, 32, 1] units with sequence length $T=10$ timesteps. Training minimizes mean absolute error on historical failure data:
{
\begin{equation}
	\mathcal{L}_{\text{RUL}} = \frac{1}{N}\sum_{i=1}^N \left|\widehat{\text{RUL}}_i - \text{RUL}_i^{\text{true}}\right|.
\end{equation}}

The model is trained separately from anomaly detection models, with batch size 32, Adam optimizer ($\alpha=10^{-3}$), and 10 epochs until convergence.

\section{LLM Prompting and Qualitative Evaluation}\label{app:llm}
\subsection{Prompt Engineering for Response Generation}

Agent C employs adaptive prompt templates based on anomaly severity. Examples:

\textbf{Low Severity ($a_{\text{fog}} < 0.6$):}
\begin{quote}
	\textit{``Equipment shows subtle sensor anomaly (severity: 0.42). Context: [temperature sensor summary]. Is this normal drift or early fault indicator? Generate: (1) likely cause; (2) continue monitoring or perform inspection?''}
\end{quote}

\textbf{High Severity ($a_{\text{fog}} > 0.8$):}
\begin{quote}
	\textit{``Critical equipment anomaly detected (severity: 0.89). Feature summary: [sensor values]. Operational impact: [cost/downtime estimates]. Recommended immediate action and required technician skills. Expected intervention timeline: [hours].''}
\end{quote}

\section{Evaluation Metrics}
\label{app:metrics}

\subsection{Anomaly Detection Metrics}

\textbf{F1-Score} (primary metric): Harmonic mean of precision and recall, computed as:
{\small 
\begin{equation}
F1 = 2 \cdot \frac{\text{Precision} \cdot \text{Recall}}{\text{Precision} + \text{Recall}} = \frac{2 \cdot \text{TP}}{2 \cdot \text{TP} + \text{FP} + \text{FN}}.
\end{equation}}
This balanced metric is critical for industrial applications where both false alarms (FP) and missed detections (FN) incur significant costs.

\textbf{Precision}: Proportion of predicted anomalies that are true anomalies, $\text{Precision} = \frac{\text{TP}}{\text{TP} + \text{FP}}$. High precision minimizes unnecessary maintenance interventions.

\textbf{Recall (Sensitivity)}: Proportion of true anomalies correctly detected, $\text{Recall} = \frac{\text{TP}}{\text{TP} + \text{FN}}$. High recall ensures critical faults are not missed.

\textbf{Accuracy}: Overall classification correctness, $\text{Accuracy} = \frac{\text{TP} + \text{TN}}{\text{Total}}$. Secondary metric due to class imbalance considerations.

\textbf{ROC-AUC}: Area under the Receiver Operating Characteristic curve, measuring probabilistic ranking quality across all threshold settings.

\subsection{RUL Prediction Metrics}

\textbf{Mean Absolute Error (MAE)}: Average absolute deviation between predicted and true RUL values:
{
\begin{equation}
\text{MAE} = \frac{1}{N}\sum_{i=1}^N |\widehat{\text{RUL}}_i - \text{RUL}_i|.
\end{equation}}

\textbf{Root Mean Square Error (RMSE)}: Square root of mean squared errors, penalizing large errors more heavily:
{
\begin{equation}
\text{RMSE} = \sqrt{\frac{1}{N}\sum_{i=1}^N (\widehat{\text{RUL}}_i - \text{RUL}_i)^2}.
\end{equation}}

\subsection{Adaptation and Efficiency Metrics}

\textbf{F1 Improvement ($\Delta$F1)}: Performance gain from initial to final iteration, $\Delta F1 = F1_{\text{final}} - F1_{\text{initial}}$, quantifying learning capability.

\textbf{Average Latency}: Mean prediction time per sample (milliseconds), critical for real-time deployment constraints.

\section{Training Configuration Details}
\label{app:training}

This appendix provides comprehensive training configurations for all models.

\subsection{Training Data Configuration}
\begin{itemize}
\item \textbf{Boiler Dataset:} 8,000 training samples, 2,000 test samples (80-20 split with stratification)
\item \textbf{Wind Turbine Dataset:} 400 training samples, 100 test samples (80-20 split)
\item \textbf{Validation Strategy:} 5-fold cross-validation on training set for hyperparameter tuning
\item \textbf{Data Augmentation:} None applied (preserves temporal integrity of sensor data)
\end{itemize}

\subsection{Anomaly Detection Models}
The anomaly detection models in Table~\ref{tab:model_training} were configured to balance detection sensitivity with computational efficiency in a streaming setting. Isolation Forest and One-Class SVM serve as unsupervised baselines that operate directly on the ISAPCA-reduced feature space, providing robust detection without labeled anomalies. The sequence models (LSTM and Transformer) capture temporal dependencies in sensor trajectories, enabling early detection of gradual degradations rather than only abrupt faults. Hyperparameters were selected through preliminary grid search to maintain stable training, avoid overfitting, and keep inference latency compatible with real-time IIoT deployment.

\begin{table}[hbt!]
  \renewcommand{\thetable}{IX}
  \centering
  \caption{Training configuration for anomaly detection models.}
  \label{tab:model_training}
  \scriptsize
  \setlength{\tabcolsep}{3.5pt}
  \renewcommand{\arraystretch}{1.0}
  \begin{tabular}{p{0.22\columnwidth} p{0.3\columnwidth} p{0.4\columnwidth}}
    \hline
    \textbf{Model} & \textbf{Hyperparam.} & \textbf{Value} \\
    \hline
    \multirow{6}{*}{\textbf{Isol. Forest}}
      & Est. ($n_\text{trees}$) & 200 \\
      & Contam. ($\rho_0$) & 0.32 \\
      & Max samples/tree & 256 \\
      & Rand. seed & 42 \\
      & Training & Unsupervised \\
    \hline
    \multirow{5}{*}{\textbf{One-Class SVM}}
      & Kernel & RBF \\
      & Gamma ($\gamma$) & $1/n_\text{feat.}$ \\
      & Nu ($\nu$) & 0.25 \\
      & Training & Unsupervised \\
    \hline
    \multirow{8}{*}{\textbf{LSTM}}
      & Arch. & [64, 32, 32, 1] units \\
      & Seq. len. & 10 timesteps \\
      & Batch & 32 \\
      & Epochs & 50 (early stop p=10) \\
      & Opt. & Adam ($\alpha=10^{-3}$, $\beta_1=0.9$) \\
      & Loss & Bin. cross-entropy \\
      & Dropout & 0.2 (post-LSTM) \\
    \hline
    \multirow{9}{*}{\textbf{Transformer}}
      & Encoder layers & 2 \\
      & Attn. heads & 4 \\
      & Emb. dim. & 128 \\
      & FF dim. & 512 \\
      & Batch & 32 \\
      & Epochs & 50 (early stop) \\
      & Opt. & Adam ($\alpha=10^{-3}$) \\
      & Loss & Bin. cross-entropy \\
    \hline
  \end{tabular}
\end{table}
\addtocounter{table}{-1}

\subsection{Hyperparameter Optimization}

All model hyperparameters are optimized through systematic grid search on validation data using 5-fold cross-validation. The search space encompasses: IF contamination $\rho \in \{0.25, 0.30, 0.32, 0.35\}$, LSTM hidden units $\in \{32, 64, 128\}$, learning rates $\in \{10^{-4}, 10^{-3}, 10^{-2}\}$, and Transformer attention heads $\in \{2, 4, 8\}$. The optimization criterion maximizes average F1-score across folds, requiring approximately 8 hours of computational time for exhaustive search across all models.

\subsection{RUL Prediction and PPO Policy Training}
The configuration in Table~\ref{tab:rul_ppo_training} couples data-driven RUL prediction with a PPO-based policy that adaptively tunes decision thresholds for anomaly management. The LSTM RUL model learns short-horizon degradation dynamics from recent sensor windows, providing a continuous estimate of remaining useful life that informs maintenance urgency. On top of this, the SEMAS PPO agent operates in the space of F1-score, precision, recall, and threshold parameters, optimizing a reward that explicitly trades off detection performance, decision stability, and intervention cost. Replay buffer size, learning rate, and clipping values were chosen to ensure stable policy improvement without oscillatory behavior, yielding a practical controller for online threshold adaptation in industrial environments.

\begin{table}[!htbp]
  \renewcommand{\thetable}{X}
  \centering
   \vspace{-0.5\baselineskip}  
  \caption{Training configuration for RUL prediction and PPO policy optimization.}
  \label{tab:rul_ppo_training}
  \scriptsize
  \setlength{\tabcolsep}{4pt}  
  \begin{tabular}{p{0.25\columnwidth}p{0.25\columnwidth}p{0.45\columnwidth}}
    \hline
    \textbf{Component} & \textbf{Parameter} & \textbf{Value/Spec.} \\
    \hline
    \multirow{6}{*}{\textbf{RUL Model}}
      & Arch. & LSTM [64,32,32,1] (ReLU) \\
      & Input & $(X_{t-10:t},\text{RUL}_t)$ \\
      & Batch & 32 \\
      & Epochs & 10 (conv. obs.) \\
      & Opt. & Adam ($\alpha=10^{-3}$) \\
      & Loss & MAE \\
    \hline
    \multirow{11}{*}{\textbf{PPO (SEMAS)}}
      & Replay & 10k $(s,a,r,s')$ \\
      & State & $[F1,P,R,w_1,w_2,\rho,\tau]$ \\
      & Action & $\Delta[w_1,w_2,\rho,\tau]$ \\
      & Reward & $\alpha F1-\beta|\Delta P-\Delta R|-\gamma L$ \\
      & Weights & $\alpha=1.0,\beta=0.1,\gamma=0.001$ \\
      & Updates & 10/iter. \\
      & Batch & 64 \\
      & LR & $3\times10^{-4}$ \\
      & Clip & 0.2 \\
      & Disc. & 0.99 \\
      & GAE $\lambda$ & 0.95 \\
    \hline
  \end{tabular}
\end{table}
\addtocounter{table}{-1}

\end{document}